\documentclass[prd, showpacs,aps,nofootinbib,floatfix,amsmath,amssymb]{revtex4}
\usepackage{graphicx}
\usepackage{soul}
\usepackage{color}
\usepackage[usenames,dvipsnames]{xcolor}
\usepackage{amsmath}
\usepackage{slashed}
\numberwithin{equation}{section}
\begin{document}

\makeatletter
\newbox\slashbox \setbox\slashbox=\hbox{$/$}
\newbox\Slashbox \setbox\Slashbox=\hbox{\large$/$}
\def\pFMslash#1{\setbox\@tempboxa=\hbox{$#1$}
  \@tempdima=0.5\wd\slashbox \advance\@tempdima 0.5\wd\@tempboxa
  \copy\slashbox \kern-\@tempdima \box\@tempboxa}
\def\pFMSlash#1{\setbox\@tempboxa=\hbox{$#1$}
  \@tempdima=0.5\wd\Slashbox \advance\@tempdima 0.5\wd\@tempboxa
  \copy\Slashbox \kern-\@tempdima \box\@tempboxa}
\def\FMslash{\protect\pFMslash}
\def\FMSlash{\protect\pFMSlash}
\def\miss#1{\ifmmode{/\mkern-11mu #1}\else{${/\mkern-11mu #1}$}\fi}

\newcommand{\psum}[1]{{\sum_{ #1}\!\!\!}'\,}
\makeatother

\title{Gauge-invariant approach to the beta function in Yang-Mills theories with universal extra dimensions}

\author{M. Huerta-Leal, H. Novales-S\'anchez, and J. J. Toscano}
\affiliation{Facultad de Ciencias F\'isico Matem\'aticas, Benem\'erita Universidad Aut\'onoma de Puebla, Apartado Postal 1152 Puebla, Puebla, M\'exico.}
\begin{abstract}
The radiative correction to beta function is comprehensively studied at the one loop level in the context of universal extra dimensions. Instead of  using cutoffs to regularize one-loop divergences, the dimensional regularization scheme is used. In this approach, large momenta  effects are removed from physical amplitudes by adjusting the parameters of the appropriate counterterms. The use of a ${\rm SU}(N)$-covariant gauge-fixing procedure to quantize the theory is stressed. One-loop contributions of Kaluza-Klein (KK) excitations are characterized by discrete KK sums and continuous momenta sums, which can diverge. Two types of ultraviolet divergences are identified, one arising from poles of the gamma function and associated with short-distance effects in the usual four-dimensional spacetime manifold, and the other emerging either from poles of the one-dimensional Epstein function or from the gamma function, and corresponding to short-distance effects in the compact manifold. We address the cases of 5 and $4+n$ dimensions ($n \geq2$) separately. In 5 dimensions the one-dimensional Epstein function is convergent, so the usual counterterm renormalizes the vacuum polarization function. For $4+n$ dimensions, the one-dimensional Epstein function is divergent, so renormalization is implemented by interactions of canonical dimension higher than four, already present in the effective theory. The polarization function is renormalized using both a mass-dependent scheme and a mass-independent scheme, with extra-dimensions effects decoupling in the former case but not in the latter. The beta function is calculated for an arbitrary number of extra dimensions. Our main result is that Yang-Mills theory remains perturbative  at the one-loop level, which is in disagreement with the results obtained in the literature by using a cutoff regulator, that suggest that Yang-Mills theory in more than four dimensions ceases to be perturbative. We emphasize the advantages of a mass-dependent scheme in this type of theories, in which decoupling is manifest.
\end{abstract}


\maketitle


\section{Introduction}
\label{Int}
The use of extra dimensions in model building started with the works by G. Nordstr\"om and T. Kaluza, who attempted to unify electromagnetism and gravity by assuming the existence of a spatial extra dimension~\cite{Nordstrom,Kaluza}. Nevertheless, it was O. Klein who realized, for the first time, that compactification could be used to explain the lack of observations of extra dimensions~\cite{Klein}. The ulterior birth of {\it string theory}, as a theory of strong interactions~\cite{Veneziano,Nielsen,KoNi1,KoNi2,Nambu,Susskind1,Susskind2}, would eventually endow great relevance to formulations of extra dimensions. The original string-theory formulation already had this ingredient, as 26 spacetime dimensions were required to ensure unitarity~\cite{Lovelace}. The introduction of fermions in string theory~\cite{Ramond}, which came along with the discovery of {\it supersymmetry}~\cite{Ramond,WeZu}, and the presence of a massless particle of spin 2~\cite{SchSch}, to be identified as the {\it graviton}, were two main elements of {\it superstring theory} that motivated its use to achieve a quantum theory of gravity, always with the complicity of extra dimensions. Remarkably, the critical dimension of superstring theory turned out to be just 10, as it was shown by J. H. Schwarz~\cite{Schwarz}. The introduction of the {\it Green-Schwarz mechanism}~\cite{GS}, to eliminate quantum anomalies arising in string theory, then triggered the {\it first superstring revolution}, during which five consistent superstring formulations were given~\cite{GS1,GHMR1,GHMR2,GHMR3}. Furthermore, a connection, through compactification, between superstring theory, featuring a 6-dimensional {\it Calabi-Yau extra-dimensional manifold}~\cite{Yau}, and 4-dimensional supersymmetry was established~\cite{CHSW}. A {\it second superstring revolution} started with the emergence of the {\it M-theory}, by E. Witten~\cite{Witten}, who showed that the five superstring  formulations known at the time are limits of this single theory, which is a unifying fundamental theory set in 11 spacetime dimensions. The existence of {\it D-branes}, proposed by J. Polchinski~\cite{Polchinski} for the sake of string duality, was a major event. It was also shown that supergravity in 11 dimensions is a low-energy limit of the $M$-theory~\cite{HoWi1,HoWi2}. The {\it ADS/CFT correspondence}, which establishes a duality of 5-dimensional theories of gravity with gauge field theories set in 4 dimensions~\cite{Maldacena}, is a quite important result with remarkable practical advantages regarding nonperturbative physics. Among the events and advances experienced by string theory throughout the years, and a plethora of papers on the matter, we wish to emphasize that its development is the one that got modern physics used to extra dimensions. \\

Considerable interest in the phenomenology of extra dimensions arose because of the works by Antoniadis, Arkani-Hamed, Dimopoulos and Dvali~\cite{A,ADD,AADD}, who, motivated by the {\it hierarchy problem}, proposed the existence of {\it large extra dimensions}, responsible for the observed weakness of the gravitational interaction, at the stunning scale of a millimeter. Shortly after, L. Randall and R. Sundrum initiated an important branch of extra-dimensional models, the so-called {\it models of warped extra dimensions}, in which the hierarchy problem was tackled by introducing a spatial extra dimension and assuming that the associated 5-dimensional spacetime is characterized by an {\it Anti-de-Sitter} structure~\cite{RS1,RS2}. The present paper is developed within another well-known extra-dimensional framework, dubbed {\it universal extra dimensions}~\cite{ACD1} (UED), proposed by Appelquist, Cheng and Dobrescu, and characterized by the assumption that every field of a given formulation depends on all the coordinates of the spacetime with extra dimensions. The work by these authors included the formulation of a field theory with the structure of the 4-dimensional Standard Model but defined, rather, on a spacetime with compact spatial extra dimensions, where all the dynamic variables are assumed to propagate, thus leading to an infinite set of {\it Kaluza-Klein} (KK) {\it modes} per each extra-dimensional field\footnote{Models of universal extra dimensions have been reviewed in Refs.~\cite{HoPr,Servant}.}. The present investigation takes place around Yang-Mills (YM) theories defined in a spacetime comprising $4+n$ dimensions, with the assumption that the $n$ extra dimensions are compact and universal. Investigations on theoretical and phenomenological aspects of extra-dimensional Yang-Mills theories have been explored and reported in Refs.~\cite{NT,FMNRT,NoTo2,LMNT1,LMNT2,GNT}. In models of universal extra dimensions, conservation of extra-dimensional momentum yields, after integrating out the extra dimensions, 4-dimensional {\it KK effective field theories} in which {\it KK parity} is preserved, with the consequence that, from the perspective of the Feynman-diagrams approach, the very first effects from the KK modes on low-energy Green's functions (and thus on low-energy observables) occur at one loop~\cite{ACD1}. Such a feature is particularly relevant in the case of physical observables and processes that, within the context of the Standard Model in 4 dimensions, can take place exclusively at loop orders. An appealing characteristic of these models is the small number of added parameters, which are a high-energy {\it compactification scale}, $R^{-1}$, and the number of extra dimensions, $n$. Moreover, universal-extra-dimensions models include {\it dark-matter candidates}~\cite{CMS,DDG,SeTa1,SeTa2}, which would be either the first {\it KK excited mode} of the photon or that corresponding to the neutrino. \\

To write down an effective Lagrangian that extends the Standard Model in the UED approach, a series of nontrivial steps must be implemented at the classical level. It is worth analyzing in some detail how this effective theory is constructed in order to have a broader understanding of its implications at the quantum level. Here, we focus on a pure (without matter fields) YM theory, whose technical details can be found in Ref.~\cite{OPYM}. At some energy scale $\Lambda$, which is assumed to be far above of the compactification scale $R^{-1}$, one proposes an effective theory governed by the extended group $ISO(1,3+n)\times SU(N,{\cal M}^{4+n})$, where $ISO(1,3+n)$ is the Poincar\' e group defined on the $(4+n)$-dimensional flat manifold ${\cal M}^{4+n}={\cal M}^4\times {\cal N}^n$, with ${\cal M}^4$ the usual spacetime manifold and ${\cal N}^n$ an Euclidean manifold. Note that $SU(N,{\cal M}^{4+n})$ and $SU(N,{\cal M}^{4})$ coincide as Lie groups, but they differ as gauge groups because they have a different number of connections. The most general effective Lagrangian that respects the $ISO(1,3+n)\times SU(N,{\cal M}^{4+n})$ symmetry can be written as: ${\cal L}_{eff}(x,\bar x)={\cal L}^{YM}_{(\mathbf{d}\leq 4+n)}(x,\bar x)+{\cal L}_{(\mathbf{d}> 4+n)}(x,\bar x)$, with $x \in {\cal M}^4$ and $\bar x \in {\cal N}^n$. The first term of this Lagrangian corresponds to the $(4+n)$-dimensional version of the usual YM theory, which contain interactions of canonical dimension less than or equal to $4+n$, while the second term includes all interactions of canonical dimension higher than $4+n$ that are compatible with these symmetries, since the theory is nonrenormalizable in the usual sense. From dimensional considerations, the interactions that appear in the ${\cal L}_{(\mathbf{d}> 4+n)}(x,\bar x)$ Lagrangian will be multiplied by inverse powers of the high energy scale $\Lambda\gg R^{-1}$, so they will be naturally suppressed with respect to those that constitute the ${\cal L}^{YM}_{(\mathbf{d}\leq 4+n)}(x,\bar x)$ Lagrangian, which do not depend on this scale. This means that this Lagrangian determines the dominant contributions to physical observables. This fact will play a central role in our quantum analysis at the one loop level. Note that the coupling constant $g_{4+n}$ that appears in this Lagrangian is dimensionful and must be rescaled to obtain the correct dimensionless YM coupling.\\

To describe physical phenomenon below the compactification scale, we need go from the $ISO(1,3+n)\times SU(N,{\cal M}^{4+n})$ description to a description based in the usual $ISO(1,3)\times SU(N,{\cal M}^{4})$ symmetry, for which it is necessary to implement a procedure of hiding of symmetry. To hide symmetry ${\rm ISO}(1,3+n)\times {\rm SU}(N,{\cal M}^{4+n})$ in symmetry ${\rm ISO}(1,3)\times {\rm SU}(N,{\cal M}^{4})$, we need to implement two canonical maps, one that transforms covariant objects of ${\rm ISO}(1,3+n)$ into covariant objects of ${\rm ISO}(1,3)$, the other that allows us to hide, through a general Fourier series, any dynamic role of subgroup ${\rm ISO}(n) \subset {\rm ISO}(1,3+n)$~\cite{OPYM}. The zero modes of such series are identified with the gauge fields and gauge parameters of the usual YM theory. To match the zero modes with those of the usual theory, the dimensionful coupling constant $g_{4+n}$ must be rescaled to identify the 4-dimensional coupling $g=g_{4+n}/(2\pi R)^{\frac{n}{2}}$. This procedure, which is qualitatively discussed in the next section, leads to an effective field theory governed by the usual $ISO(1,3)\times SU(N,{\cal M}^{4})$ groups. After compactification and integration on the $\bar x$ coordinates, the ${\cal L}^{YM}_{(\mathbf{d}\leq 4+n)}(x,\bar x)$ Lagrangian, which does not depend on the $\Lambda$ scale, unfolds into two terms, one that corresponds to the usual four-dimensional theory, ${\cal L}^{(\underline{0})}_{YM}(x)$, and the other that contains interactions between usual fields and KK excitations, ${\cal L}^{KK}_{(\mathbf{d}\leq 4)}(x)$. The important point to be emphasized is that, like ${\cal L}^{(\underline{0})}_{YM}(x)$, the ${\cal L}^{KK}_{(\mathbf{d}\leq 4)}(x)$ Lagrangian only contains interactions of canonical dimension less than or equal to four, so they cannot depend on the high energy scale $\Lambda$. This means that the ${\cal L}^{KK}_{(\mathbf{d}\leq 4)}(x)$ Lagrangian only can depend on the compactification scale $R^{-1}$ through masses of the KK excitations. This is the reason why this Lagrangian has a dominant role in phenomenological predictions, even though its effects first appear at the one-loop level. Another point worth noting is that, like ${\cal L}^{(\underline{0})}_{YM}(x)$, the interactions contained in the ${\cal L}^{KK}_{(\mathbf{d}\leq 4)}(x)$ Lagrangian are dictated by the $SU(N,{\cal M}^{4})$ gauge group, so we can expect well-behaved loop amplitudes. As far as the ${\cal L}_{(\mathbf{d}> 4+n)}(x,\bar x)$ Lagrangian is concerned, it unfolds into two infinite Lagrangians, one that depends only on zero modes fields, ${\cal L}^{(\underline{0})}_{(\mathbf{d}> 4)}(x)$, which contains all interactions of canonical dimension higher than four that respect the usual symmetry, and the other one that mixes interactions between zero modes and KK excitations. Although the ${\cal L}^{(\underline{0})}_{(\mathbf{d}> 4)}(x)$ Lagrangian can generate at the tree level the same low-energy observables that can be induced by the ${\cal L}^{KK}_{(\mathbf{d}\leq 4)}(x)$ Lagrangian at the one-loop level, its contribution is subdominant due to the suppression factor introduced by the high energy scale $\Lambda$. This means that in this type of theories, any physically interesting contribution on low-energy observables will necessarily arises from one-loop effects induced by the Lagrangians of canonical dimension less than or equal to four, namely ${\cal L}^{(\underline{0})}_{YM}(x)$ and ${\cal L}^{KK}_{(\mathbf{d}\leq 4)}(x)$. In this work, we will focus on this type of contributions.\\

 At the one-loop level, the amplitudes characterizing KK contributions are proportional to a discrete sum and a continuous sum as well, that is $\sum_{(\underline{k})}\int d^4k$, where the symbol $\sum_{(\underline{k})}$ represents a multiple sum running over discrete vectors defined on the compact manifold. The higher multiplicity of such discrete sums is the dimension of the compact manifold, $n$. Since discrete and/or continuous sums may diverge, a regularization scheme must be adopted. This is a crucial issue on which we will focus our attention. In the literature~\cite{ACD1,HTY,Kribs}, one-loop calculations of KK contributions to some observables have been presented in the cases of one and two extra dimensions using as regulator an explicit cutoff. This approach makes sense, since in an effective theory we must only include momenta below a certain energy scale and exclude those effects that are above it. Following this premise, integrals are solved using a cutoff, while only a few of the first KK modes of infinite sums are kept. The procedure used in Refs.[33,47,48] seems to be motivated by the Wilson's approach to renormalization theory, which introduce a cutoff to divide the fundamental path integral into its low-energy and high-energy parts~\cite{Collins}, although the authors only keep the low energy effects. This scheme leads, of course, to physical observables that are highly dependent on the regulator. This can invalidate the perturbation expansion even if the four-dimensional coupling constant is small. Based on these results, it is generally assumed in the literature that YM theories in more than four dimensions become strongly interacting. This result has such strong implications that it is worth reviewing more carefully. Using a cutoff as regulator to extract the low-energy effects of the theory leads to the suspicion that something could be wrong. The main goal of this work is to address this problem from a different perspective of effective theories. The first thing we must do is introduce a good regularization scheme, bearing in mind that we will face a new series of complications that are not present in conventional effective theories, which is evidenced by the presence of infinite discrete sums in the one-loop amplitudes. Although the cutoff schemes are very intuitive, they are not really good regulators because they make it difficult to keep the symmetries of the theory, especially gauge invariance, which is the essence of YM theories. Definitely, a cutoff is a bad regulator, which should not be used in theories with a high content of symmetries, like the YM theories. Although various regularization schemes are available, we know from renormalization theory that it is crucial to subtract off the infinities with a procedure that preserves all the symmetries of the original theory. In line with this observation, we will adopt the dimensional regularization scheme~\cite{BoGi,TVreg}, which has proven to be the best existing regularization scheme, mainly because it preserves the gauge and Poincar\' e symmetries. The adoption of this regularization scheme may give the impression that we are going against the genuine spirit of effective theories, since dimensional regularization keeps momenta arbitrarily high. However, as has already been shown in the literature~\cite{Weinberg,Burgess,Manohar,Wudka}, undesirable contributions, that is, the effects of very high energies, can be removed from physical observables by adjusting the parameters of appropriate counterterms. The required counterterms are already available in the effective Lagrangian, since it contains all interactions that respect the symmetries of the theory. In the case at hand, the ${\cal L}^{(\underline{0})}_{(\mathbf{d}> 4)}(x)$ Lagrangian can generate all the necessary counterterms, since it induces at the tree level all interactions between zero modes fields that are compatible with the standard symmetry $ISO(1,3)\times SU(N,{\cal M}^4)$.\\

A significant complication of field models of extra dimensions is related to the types of divergences that can arise. An infinite number of KK-mode fields, circulating in the loops of contributing Feynman diagrams, occurs and, moreover, there are two spaces, namely the usual spacetime manifold and the compact extra-dimensional manifold, so thinking of short-distance effects in both spaces makes sense. A detailed discussion on the matter, in the somewhat simpler context of extra-dimensional quantum electrodynamics, has been recently carried in Ref.~\cite{OPPRD}. As we discuss below, in the same way that there are two spaces, there are also two types of divergences, so the key to tackle the problem relies on establishing a connection between the two sorts of divergences and the two types of spaces. Such connection is already suggested by the fact that the one-loop amplitudes characterizing KK contributions are proportional, as already mentioned, to a discrete sum and a continuous sum as well, that is $\sum_{(\underline{k})}\int d^4k$. Since discrete and/or continuous sums may diverge, we simultaneously regularize them using the dimensional regularization scheme. This means that all values of both discrete and continuous momenta will be kept. As will be seen throughout the paper, the dimensional regularization scheme becomes central to our work. An important result is that the discrete sums thus regularized can naturally be expressed in terms of multidimensional Epstein functions~\cite{E1}. In this regularization scheme, divergences from continuous sums manifest as poles of the gamma function in the limit as the parameter $\epsilon=4-D$ goes to zero. We show that the divergences associated with discrete sums emerge alternately either as poles of the one-dimensional Epstein function or as poles of the gamma function. It is in this sense that we talk about the existence of two types of divergences. Since these divergences originate in discrete sums, we argue that such types of divergences correspond to short-distance effects of the KK excitations in the compact manifold, so they are genuine ultraviolet divergences that can be removed by renormalization. The KK excitations also induce divergences that emerge through poles of the gamma function, which have to do with short-distance effects in the usual spacetime manifold. Showing that both types of divergences can consistently be handled by renormalization is a main objective of the present work. In general, KK excitations can induce both types of divergences, except in the special case of only one extra dimension, since the one-dimensional Epstein function is convergent. Thus the cases $n=1$ and $n\geq 2$ are addressed separately. For $n=1$, we will show that, at the one loop level, the counterterms of the usual theory, describing the zero modes, are enough to remove ultraviolet divergences from the KK theory. However, for $n\geq 2$, short-distance effects in both the usual spacetime and the compact manifold may be present, so the usual counterterms do not suffice to remove both types of divergences. The reason is that divergences from short-distance effects in the compact manifold appear as coefficients of polynomials in the external momentum, which suggests that these effects are associated with Lagrangian terms of canonical dimension higher than four. In order to generate appropriate counterterms, renormalization is implemented by considering interactions which are nonrenormalizable, according to Dyson's criterion. As already commented, such interactions are already available in the effective KK Lagrangian, which is constituted by all the interactions governed by the symmetries of the theory.\\

{\it Asymptotic freedom} is a remarkable aspect of {\it quantum chromodynamics}, taking place in the context of the ${\rm SU}(3)_C$-invariant YM-theory description of strong interactions~\cite{GrWi,Politzer,tHooft}. This physical phenomenon, by which collisions involving hadrons display a point-like-particle behavior of constituent quarks in case of large momentum transfer, occurs in the presence of a negative {\it Callan-Symanzik beta function}~\cite{Callan,Symanzik}, $\beta$, which is identified from the ultraviolet structure of loop corrections to the gauge-field propagator. This beta function then provides, at the desired loop order, an estimation for the running {\it strong coupling constant} $\alpha_{\rm s}$, in accordance with the {\it renormalization group equation}, that is compatible with perturbation theory at sufficiently large momentum transfer $q^2$. In the one-loop approximation, $\alpha_{\rm s}(q^2)$ is expressed as
\begin{equation}
\alpha_{\rm s}(q^2)=\frac{\alpha_{\rm s}(\mu^2)}{1+\alpha_{\rm s}(\mu^2)\,\beta\log\frac{q^2}{\mu^2}},
\end{equation}
with $\mu$ denoting the {\it renormalization scale}. In the context of ${\rm SU}(3)_C$-invariant chromodynamics, the first calculation of a negative beta function, at the one-loop order, was reported in Ref.~\cite{GrWi}, whose result established asymptotic freedom of quarks and gluons. Ulterior calculations of the beta function at 2, 3, and even 4 loops were carried out afterwards~\cite{Caswell,EgTa,TVZ,LaVe,RVL,Czakon}. The presumed occurrence of KK excited modes, as a low-energy manifestation of extra dimensions, plays a role in the definition of the beta function. The efforts of the present investigation are aimed at the estimation of effects produced by the presence of universal extra dimensions on the beta function, which is done through a calculation, at one loop, of the contributions from the KK excited modes that originate in the $(4+n)$-dimensional gauge fields of the ${\rm SU}(N,{\cal M}^{4+n})$ theory.\\

A major objective of the present work is the calculation and characterization of the impact, at the one loop level, of universal extra dimensions on the beta function of ${\rm YM}$ theories. For this purpose, we take gauge symmetry, the dimensional regularization scheme, and decoupling between physical scales as guiding principles. To this aim, we quantize the KK theory by using a ${\rm SU}(N,{\cal M}^4)$-covariant gauge-fixing procedure. The Background Field Method~\cite{BFM1,BFM2} is used to fix the gauge for zero modes, which are the four-dimensional ${\rm YM}$ fields, while for KK excited modes a covariant gauge~\cite{GNT} is considered. The construction of a ${\rm SU}(N,{\cal M}^4)$-invariant quantum Lagrangian considerably simplifies the calculation of the beta function. Regarding the control of divergences, we already have emphasized the importance of using dimensional regularization to preserve the original symmetries of the theory and implementing renormalization to remove the effects of large discrete and continuous momenta. In Ref.~\cite{OPPRD}, some of us showed, for the first time, the advantages of using the dimensional regularization scheme to simultaneously regularize the continuous and discrete sums. On the other hand, we demand extra-dimensions effects to be of decoupling nature, that is, we require them to vanish from physical quantities and renormalized amplitudes at low energies, in accordance with the {\it Appelquist-Carazzone decoupling theorem}~\cite{ACDT}. The renormalized vacuum-polarization function, in our gauge-invariant quantization procedure, and the beta function shall be quantities subjected to this requirement. Decoupling of new-physics effects means that these quantities must reduce to the usual ones in the limit of a very large compactification scale. Fulfillment of this requirement depends crucially on renormalization scheme. We note that a mass-dependent scheme must be followed to achieve this goal. If these goals are achieved, we will have shown that Yang-Mills theory in more than four UED remains perturbative at the one loop level.\\

The rest of the paper has been organized as follows. In Sec.~\ref{gth}, we discuss the main features of ${\rm YM}$ theories with an arbitrary number of extra dimensions. In this section, we will focus more on the conceptual part, avoiding, as much as possible, getting into the technical details. Sec.~\ref{LC} is devoted to present the calculations needed for our study of the phenomenon of asymptotic freedom. The gauge-covariant scheme to quantize the theory is discussed to some extent. Some properties of zeta functions are presented and a number of relevant expressions for our work derived. In Sec.~\ref{AF} the beta function is derived using both a mass-independent scheme and a mass-dependent scheme. Finally, in Sec.~\ref{C} we summarize or main results and present final remarks.

\section{The extra-dimensional Yang-Mills theory and its Kaluza-Klein Lagrangian}
\label{gth}
In general, field theories are defined by {\it symmetries} and {\it dynamic variables}~\cite{Wudka}. While a variety of symmetries relevant to this purpose is available, {\it spacetime} and {\it gauge symmetries}, in particular, turn out to be essential elements for the definition of field-theory descriptions. For instance, even though most models are set under the assumption that Lorentz symmetry holds, effective field theories with the ingredient of Lorentz-invariance nonconservation, inspired by the spontaneous breaking of Lorentz symmetry in string-theory formulations~\cite{KoSa,KoPo} and by the occurrence of Lorentz violation in {\it noncommutative field theory}~\cite{CHKLO}, have been propounded~\cite{CoKo1,CoKo2,Kostelecky}. On the other hand, the choice of the gauge-symmetry group has more often become the defining trademark of field theory models. For instance, the Lagrangian terms constituting the Standard Model are determined, in part, by the gauge group ${\rm SU}(3)_C\times{\rm SU}(2)_L\times{\rm U}(1)_Y$. This is also the case of models that involve {\it left-right symmety}~\cite{PaSa,MoPa1,MoPa2,SenMo1,SenMo2,SenMo3}, which are based on SU(2)$_L\times$SU(2)$_R\times$U(1)$_{B-L}$. Moreover, the main feature of the so-called {\it 331 models}~\cite{PlPi,Frampton} is a particular gauge group, in this case SU(3)$_{C}\times$SU(3)$_L\times$U(1)$_X$. Also, theories of {\it grand unification}, based on the symmetry group SU(5), were explored and thoroughly discussed~\cite{GeGl,GQW}. The consideration of spacetime manifolds with extra dimensions has opened alternative paths to define field-theory models\footnote{Field-theory models of extra dimensions extend the 4DSM in the direction of spacetime symmetry groups.}. In the present study, we work within the framework set by $(4+n)$-dimensional YM theories, by which we refer to a replica of the ${\rm SU}(N,{\cal M}^{4})$-invariant pure-gauge theory, but with all its field content and symmetries defined on the spacetime with the $n$ extra dimensions, which we assume to be spatial-like and universal. In this section, we briefly discuss this extra-dimensional YM theory, with focus on those aspects that are relevant for the calculation of beta-function contributions that we are about to execute.
We develop our discussion in the general context of $n$ extra dimensions, which, of course, can be straightforwardly particularized to the case $n=1$.
We spare the reader from the whole bunch of specific details characterizing this formulation, and suggest Refs.~\cite{NT,FMNRT,NoTo2,LMNT1,LMNT2,OPPRD,GNT} for more detailed discussions on the matter.\\

First consider, in general, a spacetime comprising 1 time-like dimension and $3+n$ spatial-like dimensions. Assume that, at some high-energy scale (short distances), this spacetime can be characterized by a $(4+n)$-dimensional manifold ${\cal M}^{4+n}$ with metric $g_{MN}={\rm diag}(1,-1,\ldots,-1)$. Here and in what follows, capital spacetime indices, like $M,N$, take the values\footnote{Note that a convention in which the first extra dimension is labeled by $M,N=5$ has been used.} $0,1,2,3,5,\ldots,4+n$. Think of a field-theory formulation defined on this manifold and governed by the extra-dimensional Poincar\'e group ${\rm ISO}(1,3+n)$. Imagine a process in which we study nature at increasing distances, starting from the aforementioned high-energy scale. While at certain range of high-energy scales the proper field-theory description is invariant with respect to ${\rm ISO}(1,3+n)$, we assume that the afore-described process leads us to a lower-energy scale at which $n$ out of the $3+n$ spatial-like dimensions display a compact nature. It is said that these $n$ dimensions are {\it compactified}. At this energy scale, also called {\it compactification scale}, an appropriate field-theory formulation is not governed by the $(4+n)$-dimensional Poincar\'e group anymore.
Besides being a theoretical possibility, the ingredient of compactification has the practical use of explaining the absence of measurements of extra dimensions~\cite{DFK,BDDM,BKP,BBBKP,HaWe,DGKN,CTW,APS,FKKMP}. A variety of geometries, suitable for compactified extra dimensions, are available~\cite{DoPo,LNielse,CDL,MNSY,CDD}.
In general, all the symmetries and dynamic variables constituting a sensible physical description at this lower-energy scale are different.
\\

From here on, $x$ and $\bar{x}$ denote, respectively, coordinates for the 4 standard spacetime dimensions and the $n$ extra dimensions. In this context, consider any dynamic variable, generically denoted by $\chi(x,\bar{x})$, which we assume to be a tensor field with respect to the $(4+n)$-dimensional Lorentz group. Now we break ${\rm ISO}(1,3+n)$ invariance by implementing compactification, for which we assume that each extra dimension is compactified on an orbifold $S^1/Z_2$ characterized by a radius $R_j$, with $j=1,2,\ldots,n$. This compactification scheme induces periodicity properties on $\chi$, with respect to the extra-dimensional coordinates $\bar{x}$. Moreover, it allows for the assignment to $\chi$ of definite-parity properties, with respect to reflections $\bar{x}\to-\bar{x}$. The field $\chi$ is then expanded in terms of a {\it complete set of orthogonal functions} $\{ f^{(\underline{k})}_{\rm E}(\bar{x}), f^{(\underline{k})}_{\rm O}(\bar{x}) \}$, which exclusively depend on extra-dimensional coordinates $\bar{x}$. Such an expansion runs over the multi-index $(\underline{k})=(k_1,k_2,\ldots,k_n)$, where any $k_j$ is an integer number. Furthermore, the  labels ``E'' and ``O'' mean that the corresponding function is even or odd under $\bar{x}\to-\bar{x}$. Once this expansion has been implemented on every field $\chi$, the extra-dimensional coordinates $\bar{x}$ no longer label degrees of freedom, which are now characterized by the {\it KK index} $(\underline{k})$. Each function $f^{(\underline{k})}_{\rm E}$ or $f^{(\underline{k})}_{\rm O}$, in the $\chi$ expansion, lies multiplied by a coefficient $\chi^{(\underline{k})}_{\rm E}(x)$ or $\chi^{(\underline{k})}_{\rm O}(x)$, respectively. These fields, which depend only on the four-dimensional coordinates $x$ of the non-compact spacetime dimensions, are the new four-dimensional dynamic variables, the KK modes, suitable for the physical description after compactification. Assume that a constant function, $f^{(\underline{0})}$, belongs to $\{ f^{(\underline{k})}_{\rm E}, f^{(\underline{k})}_{\rm O} \}$. This function, trivially even under $\bar{x}\to-\bar{x}$, comes with a 4-dimensional field $\chi^{(\underline{0})}(x)$.
Fields $\chi^{(\underline{0})}$, known as KK zero modes, are identified as the dynamic variables that constitute the low-energy description. The fields $\chi^{(\underline{k})}$, with $(\underline{k})\ne(\underline{0})$, are known as KK excited modes, and are interpreted as degrees of freedom that reflect the presence of extra dimensions from a four-dimensional effective-theory viewpoint. Thus, only $\bar{x}$-even extra-dimensional fields $\chi$ yield low-energy dynamic variables. \\

The specific set $\{ f^{(\underline{k})}_{\rm E}, f^{(\underline{k})}_{\rm O} \}$ is determined, in part, by the geometry of the extra dimensions, but an extra-dimensional observable is also required to this end. This is the case of the {\it Casimir invariants} of ${\rm ISO}(n)$, among which we choose $\bar{P}^2$, with $\bar{P}$ the momentum operator along the extra dimensions.
Being a hermitian operator, $\bar{P}^2$ has an associated orthogonal set of eigenkets $\{ | \bar{p}^{(k)} \rangle \}$, with real eigenvalues $(\bar{p}^{(\underline{k})})^2=\bar{p}^{(\underline{k})}\cdot\bar{p}^{(\underline{k})}$. The $\bar{P}^2$ eigenkets then define $\{ f^{(\underline{k})}_{\rm E}, f^{((\underline{k})}_{\rm O} \}$ from the wave-function relations $f^{(\underline{k})}_{\rm E,O}=\langle  \bar{x} | \bar{p}^{(\underline{k})} \rangle$.
Using such relations, together with appropriate boundary conditions, $f^{(\underline{k})}_{\rm E}$ and $f^{(\underline{k})}_{\rm O}$ are determined to be normalized trigonometric functions, so that the field $\chi$ is Fourier expanded,
with the following two disjoint cases:
\\ \\
{\it Even parity:}
\begin{eqnarray}
\chi(x,\bar{x})=\frac{1}{\sqrt{(2\pi)^n{\cal R}}}\,\,\chi^{(\underline{0})}(x)
+\sum_{(\underline{k})}\sqrt{\frac{2}{(2\pi)^n{\cal R}}}\,\,\chi_{\rm E}^{(\underline{k})}(x)\cos\{\bar{p}^{(\underline{k})}\cdot\bar{x}\},
\label{efields}
\end{eqnarray}
\\
{\it Odd parity:}
\begin{equation}
\chi(x,\bar{x})=\sum_{(\underline{k})}\sqrt{\frac{2}{(2\pi)^n{\cal R}}}\,\,\chi_{\rm O}^{(\underline{k})}(x)\sin\{\bar{p}^{(\underline{k})}\cdot\bar{x}\}.
\label{ofields}
\end{equation}
In these equations, we denoted ${\cal R}=R_1\,R_2\cdots R_n$.
We have defined discrete extra-dimensional momenta $\bar{p}^{(\underline{k})}=(k_1/R_1,k_2/R_2,\ldots,k_n/R_n)$ as well. The symbol $\sum_{(\underline{k})}=\sum_{k_1}\sum_{k_2}\cdots\sum_{k_n}$ represents a multiple sum that runs over every discrete vector $(\underline{k})$ labeling an independent field $\chi^{(\underline{k})}$, with the additional restriction that $(\underline{k})\ne(\underline{0})=(0,0,\ldots,0)$. That is, this notation summarizes a total of $ 2^{n}-1 $ different series as follows:
\begin{align}
&\sum_{(\underline{k})}T^{(\underline{k})}  =  \sum_{k_{1}=1}^{\infty}T^{(k_{1},0,\ldots,0)}+\ldots+ \sum_{k_{n}=1}^{\infty}T^{(0,\ldots,k_n)} \nonumber \\
&+\sum_{k_{1},k_{2}=1}^{\infty}T^{(k_{1},k_{2},0,\ldots,0)}+\ldots+ \sum_{k_{n-1},k_{n}=1}^{\infty}T^{(0,\ldots,0,k_{n-1},k_n)} \nonumber \\
&\vdots \nonumber \\
 &+\sum_{k_{1},\ldots, k_{n}=1}^{\infty}T^{(k_{1},\ldots, k_{n})}\ .\label{SD}
\end{align}
Whereas positions of Fourier indices in the entries of $(\underline{k})$ are not relevant, the number of occupied entries makes a difference. So, in practice, one can use the following definition
\begin{equation}
\label{A2}
\sum_{(\underline{k})}=\sum^n_{l=1}\left(\begin{array}{ccc}
n \\
l
\end{array}\right)\sum^\infty_{k_1=1}\cdots \sum^\infty_{k_l=1}\, .
\end{equation}
Note that such an effective theory, often referred to as {\it KK theory}, is defined only in 4 dimensions of spacetime: once the extra dimensions have been compactified and the dynamic variables Fourier expanded, the whole dependence on extra dimensional coordinates in the action $S^{\rm YM}_{4+n}=\int d^{4+n}x\,{\cal L}^{\rm YM}_{4+n}(x,\bar{x})$ lies within trigonometric functions, which can be straightforwardly integrated out, leading to a Lagrangian
\begin{equation}
{\cal L}^{\rm YM}_{\rm KK}(x)=\int d^n\bar{x}\,{\cal L}^{\rm YM}_{4+n}(x,\bar{x})
\label{LYMKK}
\end{equation}
defined in 4 spacetime dimensions.\\

We start by considering a YM theory set on a $(4+n)$-dimensional spacetime with those features previously described and which is governed by the extended groups ${\rm ISO(1,3+n)}\times {\rm SU}(N,{\cal M}^{4+n})$.
The gauge-symmetry group introduces $N^2-1$ connections, denoted as ${\cal A}^a_M(x,\bar{x})$, where $a=1,2,\ldots,N^2-1$ is the gauge index.
The theory to be addressed is then given by the Lagrangian term
\begin{equation}
\label{CL4n}
{\cal L}^{\rm YM}_{4+n}=-\frac{1}{4}{\cal F}^a_{MN}{\cal F}^{aMN}+{\cal L}_{(\mathbf{d}>4+n)}\left({\cal F}, {\cal D}{\cal F}\right)\, ,
\end{equation}
where ${\cal L}_{(\mathbf{d}>4+n)}\left({\cal F}, {\cal DF}\right)$ includes all interactions which have canonical dimension higher than $4+n$ and which are compatible with the ${\rm ISO(1,3+n)}\times {\rm SU}(N,{\cal M}^{4+n})$ symmetry. This must be so because the theory is nonrenormalizable in the usual sense. This effective Lagrangian is given in terms of the ${\rm SU}(N,{\cal M}^{4+n})$ YM curvature components ${\cal F}^a_{MN}$ and its derivatives, defined as usual~\cite{PeSch}:
\begin{subequations}
\begin{align}
{\cal F}^a_{MN}&=\partial_M{\cal A}^a_N-\partial_N{\cal A}^a_M+g_{4+n}\,f^{abc}{\cal A}^b_M{\cal A}^c_N\, , \\
{\cal D}^{ab}_M&=\delta^{ab}\partial_M-g_{4+n}f^{abc}{\cal A}^c_M\, ,
\end{align}
\end{subequations}
with $f^{abc}$ the {\it structure constants} of the group. The ${\rm SU}(N,{\cal M}^{4+n})$ {\it coupling constant}, $g_{4+n}$,
is dimensionful, with units $({\rm mass})^{-n/2}$.
\\

Once defined the Lagrangian term ${\cal L}^{\rm YM}_{4+n}$, we implement compactification through a couple of canonical transformations to go from the $(4+n)$-dimensional perspective to the KK effective theory, set in 4 spacetime dimensions. Due to compactification, ${\cal A}^a_M$, which at first was a $(4+n)$-vector of ${\rm SO}(1,3+n)$, is split into the ${\rm SO}(1,3)$ 4-vector ${\cal A}^a_\mu$, and the set of ${\rm SO}(1,3)$ scalar fields $\{{\cal A}^a_5,{\cal A}^a_6,\ldots,{\cal A}^a_{4+n}\}$. From now on, we utilize greek indices like $\mu,\nu=0,1,2,3$ to denote 4-dimensional Lorentz indices and use indices $\bar{\mu},\bar{\nu}=5,6,\ldots,4+n$ to label extra-dimensions coordinates. The implementation of the afore-alluded splitting is a canonical transformation that maps covariant objects of ${\rm SO}(1,3+n)$ into covariant objects of ${\rm SO}(1,3)$~\cite{LMNT1,LMNT2}. In order to land on a KK {\it effective Lagrangian} consistently comprising the low-energy theory, namely the YM theory in 4 dimensions, we assume that ${\cal A}^a_\mu$ is even with respect to $\bar{x}\to-\bar{x}$, but the definite parity of the scalar fields ${\cal A}^a_{\bar{\mu}}$ under such a transformation is taken to be odd.
Eqs.~(\ref{efields}) and (\ref{ofields}) embody a second canonical transformation~\cite{LMNT1,LMNT2} which, after implementation, yields sets of KK modes, recognized as dynamic variables of the KK Lagrangian. Furthermore, as a byproduct of the last canonical transformation, the dimensionless quantity $g=g_{4+n}/\sqrt{(2\pi{\cal R})^n}$ is identified as the ${\rm SU}(N,{\cal M}^4)$ coupling constant. The whole set of KK modes, together with the two canonical maps generating them, is illustrated in Eq.~(\ref{canmaps}):
\begin{equation}
\begin{array}{rl}
{\cal A}^a_M(x,\bar{x})&\mapsto
\left\{
\begin{array}{l}
{\cal A}^a_\mu(x,\bar{x})\mapsto
A^{(\underline{0})a}_\mu(x),\,
A^{(\underline{k})a}_\mu(x)
\vspace{0.3cm} \\
{\cal A}^a_{\bar{\mu}}(x,\bar{x})\mapsto A^{(\underline{k})a}_{\bar{\mu}}(x)
\end{array}
\right.
\end{array}
\label{canmaps}
\end{equation}
\\

After usage of the canonical maps, and subsequent straightforward integration of the extra dimensions in the action, the 4-dimensional KK Lagrangian term ${\cal L}^{\rm YM}_{\rm KK}=\int d^n\bar{x}\,{\cal L}^{\rm YM}_{4+n}$ arises. The effective-theory description provided by ${\cal L}^{\rm YM}_{\rm KK}$ is characterized by low-energy symmetries, among which 4-dimensional Poincar\'e invariance is central. With respect to the Lorentz group ${\rm SO}(1,3)$, the KK fields $A^{(\underline{0})a}_\mu$ and $A^{(\underline{k})a}_\mu$ are 4-vectors, whereas the $A^{(\underline{k})a}_{\bar{\mu}}$
are scalars. About gauge symmetry, the effectuation of compactification entails the occurrence of {\it hidden symmetries}~\cite{LMNT1}. Originally characterized by the gauge group ${\rm SU}(N,{\cal M}^{4+n})$, set on $4+n$ spacetime dimensions, the $(4+n)$-dimensional Yang-Mills theory has been mapped into a KK theory that manifests gauge invariance corresponding to the low-energy group ${\rm SU}(N,{\cal M}^{4})$, defined on 4 spacetime dimensions. Collaterally, the gauge transformations of the $(4+n)$-dimensional connections ${\cal A}^a_M$ split into two disjoint sets of 4-dimensional gauge transformations~\cite{NT,FMNRT,NoTo2,LMNT1,LMNT2,GNT}: {\it standard gauge transformations}, which constitute the gauge group ${\rm SU}(N,{\cal M}^{4})$ and with respect to which KK zero mode $A^{(\underline{0})a}_\mu$ behaves as a gauge field; {\it nonstandard gauge transformations}, under which KK excited modes $A^{(\underline{k})a}_\mu$ are sort of like gauge fields, in the sense that they follow a transformation that is reminiscent of a gauge transformation, but which does not correspond to ${\rm SU}(N,{\cal M}^{4})$.
Furthermore, let us remark that KK excited modes $A^{(\underline{k})a}_{\mu}$ are not connections of ${\rm SU}(N,{\cal M}^{4})$, but, rather, they transform as matter fields, in the adjoint representation of this group~\cite{NT,FMNRT,NoTo2,LMNT1,LMNT2,GNT}. Hence gauge symmetry governing the KK effective theory does not forbid the presence of mass terms for vector KK-excited-mode fields $A^{(\underline{k})a}_\mu$. This is to be contrasted with the situation of KK zero modes $A^{(\underline{0})a}_\mu$, which, being 4-dimensional gauge fields, are restricted to be massless. All scalar KK modes, on the other hand, transform as matter fields with respect to both sets of gauge transformations; the scalar fields $A^{(\underline{k})a}_{\bar{\mu}}$ are, in spite of their gauge origin, matter fields under ${\rm SU}(N,{\cal M}^{4})$, which opens the possibility for they to become massive.
\\

A remarkable outcome of compactification is the occurrence of mass terms for the whole set of KK excited modes, which we refer to as the {\it KK mass-generating mechanism}, or {\it KK mechanism} for short. To illustrate more explicitly the consequences of this mechanism, we present the effective Lagrangian that results after compactification of the $(4+n)$-dimensional version of the pure Yang-Mills theory. So, after integrating over the extra coordinates in (\ref{LYMKK}), we have an effective Lagrangian given by
\begin{equation}
\label{CL}
{\cal L}^{\rm YM}_{\rm KK}={\cal L}^{\rm YM}_{{\rm KK}(\mathbf{d}\leq 4)}+{\cal L}_{(\mathbf{d}>4)}\, ,
\end{equation}
where ${\cal L}^{\rm YM}_{{\rm KK}(\mathbf{d}\leq 4)}$ contains only interactions of canonical dimension less than or equal to 4, which emerge from compactification of the $(4+n)$-dimensional version of the YM theory. This Lagrangian can conveniently be written as follows:
\begin{equation}
{\cal L}^{\rm YM}_{{\rm KK}(\mathbf{d}\leq 4)}= {\cal L}^{\rm YM}_{\textrm{v-v}}+ {\cal L}^{\rm YM}_{\textrm{v-s}}+ {\cal L}^{\rm YM}_{\textrm{s-s}}
\end{equation}
where
\begin{subequations}
\begin{align}
\label{ELVV}
{\cal L}^{\rm YM}_{\textrm{v-v}}&=-\frac{1}{4}{\cal F}^{(\underline{0})a}_{\mu \nu}{\cal F}^{(\underline{0})\mu \nu}_a -\frac{1}{4}
\sum_{(\underline{k})}{\cal F}^{(\underline{k})a}_{\mu \nu}{\cal F}^{(\underline{k})\mu \nu}_a\, , \\
\label{ELVS}
{\cal L}^{\rm YM}_{\textrm{v-s}}&=\frac{1}{2}
\sum_{(\underline{k})}{\cal F}^{(\underline{k})a}_{\mu \bar \nu}(x){\cal F}^{(\underline{k})a\, \mu}\hspace{0.00001cm}_{\bar \nu}(x)\, , \\
\label{ELSS}
{\cal L}^{\rm YM}_{\textrm{s-s}}&=-\frac{1}{4}{\cal F}^{(\underline{0})a}_{\bar \mu \bar \nu}{\cal F}^{(\underline{0})\bar \mu \bar \nu}_a-\frac{1}{4}
\sum_{(\underline{k})}{\cal F}^{(\underline{k})a}_{\bar \mu \bar \nu}{\cal F}^{(\underline{k})\bar \mu \bar \nu}_a\, .
\end{align}
\end{subequations}
The Lagrangian ${\cal L}_{(\mathbf{d}>4)}$ in (\ref{CL}) arise from compactification of ${\cal L}_{(\mathbf{d}>4+n)}$ in (\ref{CL4n}), which contains all interactions of canonical dimension higher than 4, which are compatible with the standard symmetry ${\rm ISO}(1,3)\times {\rm SU}(1,{\cal M}^4)$. In the above expressions, the curvature components $\{{\cal F}^{(\underline{0})a}_{\mu \nu}(x), {\cal F}^{(\underline{k})a}_{\mu \nu}(x)\}$ are given by
\begin{subequations}
\begin{align}
\label{CVV0}
{\cal F}^{(\underline{0})a}_{\mu \nu}&=F^{(\underline{0})a}_{\mu \nu}+gf^{abc}\sum_{(\underline{k})}A^{(\underline{k})b}_\mu A^{(\underline{k})c}_\nu \, ,\\
\label{CVVm}
{\cal F}^{(\underline{k})a}_{\mu \nu}&={\cal D}^{(\underline{0})ab}_\mu A^{(\underline{k})b}_\nu-{\cal D}^{(\underline{0})ab}_\nu A^{(\underline{k})b}_\mu\nonumber \\
&+gf^{abc}\sum_{(\underline{rs})}\Delta_{(\underline{krs})}A^{(\underline{r})b}_\mu A^{(\underline{s})c}_\nu\, ,
\end{align}
\end{subequations}
where
\begin{equation}
\label{CVV00}
F^{(\underline{0})a}_{\mu \nu}=\partial_\mu A^{(\underline{0})a}_\nu-\partial_\nu A^{(\underline{0})a}_\mu+gf^{abc}A^{(\underline{0})b}_\mu A^{(\underline{0})c}_\nu\, ,
\end{equation}
are the curvature components associated with the standard gauge group ${\rm SU}(N,{\cal M}^4)$. Note that the curvature components $\{{\cal F}^{(\underline{0})a}_{\mu \nu}(x), {\cal F}^{(\underline{k})a}_{\mu \nu}(x)\}$ transform in the adjoint representation of this group. Then, the Lagrangian (\ref{ELVV}) contains the Yang-Mills term associated with the standard ${\rm SU}(N,{\cal M}^4)$ group plus terms involving interactions among connection components and matter fields. Note that ${\cal F}^{(\underline{k})a}_{\mu \nu}(x)$ contains the kinetic terms for the matter fields. As far as the curvatures $\{{\cal F}^{(\underline{k})a}_{\mu \bar \nu}, {\cal F}^{(\underline{k})a}_{\bar \mu \bar \nu}, {\cal F}^{(\underline{0})a}_{\bar \mu \bar \nu} \}$ are concerned, which transform in the adjoint representation of the gauge group ${\rm SU}(N,{\cal M}^4)$, they are given by:
\begin{subequations}
\begin{align}
\label{CVF}
{\cal F}^{(\underline{k})a}_{\mu \bar \nu}&={\cal D}^{(\underline{0})ab}_\mu A^{(\underline{k})b}_{\bar \nu}+p^{(\underline{k})}_{\bar \nu}A^{(\underline{k})a}_{ \mu}\nonumber \\
&+gf^{abc}\sum_{(\underline{rs})}\Delta'_{(\underline{rsk})}A^{(\underline{r})b}_\mu(x)A^{(\underline{s})c}_{\bar \nu}\, , \\
\label{CSF}
{\cal F}^{(\underline{k})a}_{\bar \mu \bar \nu}&=p^{(\underline{k})}_{\bar \mu}A^{(\underline{k})a}_{\bar \nu}-p^{(\underline{k})}_{\bar \nu}A^{(\underline{k})a}_{\bar \mu}\nonumber \\
&+gf^{abc}\sum_{(\underline{rs})}\Delta'_{(\underline{rsk})}A^{(\underline{r})b}_{\bar \mu}A^{(\underline{s})c}_{\bar \nu}\, , \\
\label{CSF0}
{\cal F}^{(\underline{0})a}_{\bar \mu \bar \nu}&=gf^{abc}\sum_{(\underline{k})}A^{(\underline{k})b}_{\bar \mu}(x)A^{(\underline{k})c}_{\bar \nu}\, .
\end{align}
\end{subequations}
In the above expressions, ${\cal D}^{(\underline{0})ab}_\mu$ is the covariant derivative in the adjoint representation of ${\rm SU}(N,{\cal M}^4)$. In addition, the symbols $\Delta_{(\underline{rks})}$ and $\Delta'_{(\underline{rks})}$ are given by

\begin{subequations}
\begin{align}
\Delta_{(\underline{rks})}&=\frac{1}{f^{(\underline{0})}_E}\int_{0}^{2\pi R_n}\ldots\int_{0}^{2\pi R_1}\mathrm{d}^{n}\overline{x} f^{(\underline{k})}_{E}( \bar{x})f^{(\underline{s})}_{E}( \bar{x})f^{(\underline{r})}_{E}( \bar{x})\, ,\\
\Delta'_{(\underline{rks})}&=\frac{1}{f^{(\underline{0})}_E}\int_{0}^{2\pi R_n}\ldots\int_{0}^{2\pi R_1}\mathrm{d}^{n}\overline{x} f^{(\underline{k})}_{O}(\bar{x})f^{(\underline{s})}_{O}\bar{x})f^{(\underline{r})}_{E}(\bar{x})\, .
\end{align}
\end{subequations}
Note that the Lagrangians (\ref{ELVS}) and (\ref{ELSS}) correspond to a scalar kinetic sector and to a scalar potential, respectively. This means that  masses for the gauge, $A^{(\underline{k})a}_\mu(x)$, and scalar, $A^{(\underline{k})a}_{\bar \mu}(x)$, fields can arise from the Lagrangians (\ref{ELVS}) and (\ref{ELSS}), respectively.\\

Any KK excited mode $\chi^{(\underline{k})}$, labeled by a specific multi-index $(\underline{k})$, acquires a {\it KK mass},
\begin{equation}
m_{(\underline{k})}=\sqrt{\left(\frac{k_1}{R_1}\right)^2+\left(\frac{k_2}{R_2}\right)^2+\cdots+\left(\frac{k_n}{R_n}\right)^2}.
\end{equation}
Mass terms for vector KK fields $A^{(\underline{k})a}_\mu$ emerge in a straightforward manner from Eqs.(\ref{ELVS}) and (\ref{CVF}),
which contrasts with the case of scalar KK excited modes belonging to the set $\{A^{(\underline{k})a}_5,A^{(\underline{k})a}_6,\ldots,A^{(\underline{k})a}_{4+n}\}$, with fixed KK index $(\underline{k})$, since mixings among all the fields of such a set take place, as it follows from Eqs.(\ref{ELSS}) and (\ref{CSF}).
The mixing for this set of scalar KK modes is given by the real and symmetric mixing matrix ${\cal M}^{(\underline{k})}$, with entries ${\cal M}^{(\underline{k})}_{\bar{\mu}\bar{\nu}}=m_{(\underline{k})}^2\,\delta_{\bar{\mu}\bar{\nu}}-\overline{p}^{(\underline{k})}_{\bar{\mu}}\overline{p}^{(\underline{k})}_{\bar{\nu}}$. Things can be conveniently arranged so that, denoting the orthogonal-diagonalization matrix of ${\cal M}^{(\underline{k})}$ by $R^{(\underline{k})}$, the diagonalization\footnote{In this equation, the repeated index $\bar{\mu}$ is not summed.} $(R^{(\underline{k}){\rm T}}{\cal M}^{(\underline{k})}R^{(\underline{k})})_{\bar{\mu}\bar{\nu}}=m_{(\underline{k})}^2\delta_{\bar{\mu}\bar{\nu}}(1-\delta_{\bar{\mu},4+n})$ can be executed. Note that the eigenvalues of $R^{(\underline{k}){\rm T}}{\cal M}^{(\underline{k})}R^{(\underline{k})}$ are $m_{(\underline{k})}^2$, except for that corresponding to $\bar{\mu}=\bar{\nu}=4+n$, which is 0. The null eigenvalue implies the presence of massless scalar KK excited modes, which we denote as $A^{(\underline{k})a}_{\rm G}$, and which turn out to be kind of pseudo-Goldstone bosons, in the sense that a nonstandard gauge transformation that eliminates them from the theory exists, indicating that such fields represent unphysical degrees of freedom. After the change of basis, induced by diagonalization, the resulting mass-eigenfields basis involves, for any fixed KK index $(\underline{k})$, the set of scalar fields $\{ A'^{(\underline{k})a}_1,A'^{(\underline{k})a}_2,\ldots,A'^{(\underline{k})a}_{n-1} \}$, all of them with mass $m_{(\underline{k})}$, and the aforementioned pseudo-Goldstone bosons $A^{(\underline{k})a}_{\rm G}$. Such a diagonalization, with the associated set of resulting fields, is illustrated in Eq.~(\ref{diagYM}):
\begin{equation}
\begin{array}{c}
\big\{A^{(\underline{k})a}_{\bar{\mu}}\big\}_{\bar{\mu}=5}^{4+n}\,
\mapsto
A^{(\underline{k})a}_{\rm G},\,\big\{A'^{(\underline{k})a}_{\bar{n}}\big\}_{\bar{n}=1}^{n-1}
\end{array}
\label{diagYM}
\end{equation}

The KK-mechanism procedure bears features that evoke the {\it Englert-Higgs mechanism} (EHM)~\cite{EnBr,PWHiggs1,PWHiggs2}, responsible for mass generation in the Standard Model. A gauge-invariant scalar potential with degenerate minima, which can be characterized by the set of points constituting a hypersphere with radius determined by some {\it vacuum expectation value}, is the starting point of the EHM. The hypersphere points are connected to each other by gauge symmetry associated to some group $G$, of dimension $d_G$, so they represent physically equivalent vacuum states. To pick one of such minima, a specific constant vector, associated to a particular point on the hypersphere, is taken. Such a choice induces a map $G\mapsto H$ that breaks the gauge group $G$ down into one of its subgroups $H\subset G$, of dimension $d_H$. This procedure breaks $d_G-d_H$ generators of $G$, thus leaving $d_H$ unbroken generators. Any gauge field pointing towards the direction defined by a broken generator becomes massive, which yields the emergence of an associated pseudo-Goldstone boson. Hence the resulting set of fields involves $d_G-d_H$ massive gauge fields and the same number of pseudo-Goldstone bosons. On the other hand, the $d_H$ gauge fields pointing along directions corresponding to unbroken generators remain massless and are the connections of the gauge subgroup $H$, which governs the resultant theory. So, the remaining $d_H$ unbroken generators define the Lie algebra of $H$. Regarding the KK mechanism, note that the complete set of orthogonal functions $\{ f^{(\underline{k})}_{\rm E}, f^{((\underline{k})}_{\rm O} \}$ is not unique. In order to pick a particular set, an extra-dimensional observable, namely the ${\rm ISO}(n)$ Casimir invariant $\bar{P}^2$, was utilized, though other options, yielding different sets $\{ f^{(\underline{k})}_{\rm E}, f^{((\underline{k})}_{\rm O} \}$, are available. The definition of a such a set determines a canonical transformation that maps the extra-dimensional fields into the 4-dimensional KK modes, thus defining a theory governed by 4-dimensional Poincar\'e invariance. In other words, the map ${\rm ISO}(1,3+n)\to{\rm ISO}(1,3)$ takes place. Furthermore, while the extra-dimensional theory is invariant with respect to some gauge group defined on the spacetime with extra dimensions, after this map the resulting theory is manifestly governed by a gauge group characterized by the same generators, though defined in 4 dimensions. Consider a connection of the gauge group in extra dimensions and assume that it has been mapped into its set of KK modes. The corresponding KK zero mode points along the direction of the constant function $f^{(\underline{0})}=\langle \bar{x} | \bar{p}^{(\underline{0})} \rangle$, determined by the $\bar{P}^2$ eigenket $| \bar{p}^{(\underline{0})} \rangle$. The zero mode remains massless and transforms as a gauge field with respect to the 4-dimensional gauge group, which resembles what happens with the gauge fields pointing towards the directions associated to unbroken generators in the EHM. Moreover, the remaining $2^n-1$ eigenkets $| \bar{p}^{(\underline{k})} \rangle$, with $(\underline{k})\ne(\underline{0})$, are analogues of the broken gauge-group generators from the EHM, in the sense that they define independent directions $f^{(\underline{k})}_{\rm O}$ and $f^{(\underline{k})}_{\rm E}$ along which vector fields with masses acquired by the KK mechanism are directed, with the presence of the same number of associated pseudo-Goldstone bosons. It is worth emphasizing that, in contrast with the case of the EHM, the KK mechanism does not involve broken gauge generators, since the extra-dimensional and the 4-dimensional gauge groups share the same generators.
\\

\section{The one-loop calculation}
\label{LC}The purpose of this section is to present the calculation of the one-loop contribution of the KK modes to the tensor polarization associated with the zero-mode gauge field $A^{(\underline{0})a}_\mu$.

\subsection{The gauge-fixed renormalized action}
Our goal is the calculation, at one loop, of the beta function associated to the gauge group ${\rm SU}(N,{\cal M}^4)$, in the context of the KK theory discussed in the previous section. To this aim, we are interested in those couplings that contribute at one loop to low-energy Green's functions, characterized by Feynman diagrams in which the external fields are exclusively zero-mode gauge fields $A^{(\underline{0})a}_\mu$. Such Lagrangian terms include KK excited-mode vector fields $A^{(\underline{k})a}_\mu$ and scalar fields $A'^{(\underline{k})a}_{\bar n}$, but note that contributions from the zero mode $A^{(\underline{0})a}_\mu$ must be considered as well~\cite{PeSch}. So, an effective action $\Gamma[A^{(\underline{0})}]$, which results from integrating out quantum fluctuations of $A^{(\underline{0})a}_\mu$ and KK excited-mode fields, is defined and calculated. \\

We address gauge fixing in the KK theory within the framework of the {\it Becchi-Rouet-Stora-Tyutin} (BRST) formalism. The extended-action proper solution for the 4-dimensional gauge group ${\rm SU}(N)$ has been discussed in detail in Ref.~\cite{GPS}, while a generalization to extra dimensions and the corresponding KK theory are found in Refs.~\cite{NT,NoTo2,GNT}. In this approach, gauge fixing in the Standard Model in extra dimensions has been discussed in Refs.~\cite{CGNT,MNT}. The implementation of these techniques to the $(4+n)$-dimensional YM theory and its KK effective description yields the quantum Lagrangian
\begin{equation}
\label{QL}
{\cal L}^{\rm YM}_{\rm QKK}={\cal L}^{\rm YM}_{\rm KK}+{\cal L}^{\rm GF}_{\rm KK}+{\cal L}^{\rm G}_{\rm KK}+{\cal L}^{(\underline{0}){\rm YM}}_{\rm c.t.}+{\cal L}^{\rm KK}_{\rm c.t.}\, ,
\end{equation}
where ${\cal L}^{\rm YM}_{\rm KK}$ is the classical Lagrangian, Eq.~(\ref{CL}), ${\cal L}^{\rm GF}_{\rm KK}$ is the gauge-fixing term, and ${\cal L}^{\rm G}_{\rm KK}$ is the corresponding ghost-antighost sector. In addition, ${\cal L}^{(\underline{0}){\rm YM}}_{\rm c.t.}$ is the usual YM counterterm, whereas ${\cal L}^{\rm KK}_{\rm c.t.}$ contains the remaining counterterms of the effective theory.\\

\subsection{Gauge fixing in the Kaluza-Klein theory}
Field formulations aimed at furnishing sensible quantum descriptions of nature are usually built on the grounds of gauge symmetry. The essence of gauge symmetry resides in the presence of more degrees of freedom than those strictly required by some given system for its description~\cite{HenTe}. Gauge transformations link a whole family of different mathematical configurations which, in order for gauge symmetry to make physical sense, must lead to the exact same physical results. In other words, any observable intended to be genuinely physical must be {\it gauge independent}. Even though gauge symmetry is a main element for the definition of field theories, it turns out that quantization requires {\it gauge fixing} to be carried out, which means to choose a specific gauge, thus resulting in a formulation that is not gauge invariant anymore. \\

Being associated to local symmetry groups, gauge transformations are defined by functions, known as {\it gauge parameters}, which depend on spacetime coordinates. The selection of a set of specific spacetime-dependent functions to play the role of gauge parameters fixes the gauge, establishing a particular gauge configuration. A systematic path to pick a gauge, among the so-called {\it linear gauges}, was developed long ago by the authors of Ref.~\cite{FLS}. In their approach, gauge fixing is parametrized by a {\it gauge-fixing parameter}, usually denoted as $\xi$, whose different values correspond to different gauges. In such an approach, the {\it Landau gauge}, $\xi=0$, and the {\it Feynman-'t Hooft gauge}, $\xi=1$, are commonly utilized. When massive gauge fields are present, a customary choice is the {\it unitary gauge}, which, in this scheme, is obtained by taking the limit as $\xi\to\infty$. \\

The {\it field-antifield formalism} and the BRST symmetry constitute an efficacious mean through which the quantization of gauge systems can be achieved~\cite{GPS,BaVi1,BaVi2,BaVi3,BaVi4,BaVi5,BRS1,BRS2,Tyutin}. In this framework, the field content defining some gauge theory gets systematically extended. First, a set of {\it ghost} and {\it antighost fields} is added to the theory; more precisely, per each gauge parameter participating in the theory, a ghost-antighost pair is introduced. Also, a set of {\it auxiliary fields} is included. Then, a further enlargement of the field content takes place by the incorporation of antifields, one per each field already defined. Moreover, a {\it symplectic structure}, known as {\it the antibracket} is defined, with each field-antifield pair being {\it canonical conjugate variables}. The resultant increased set of fields is then understood to define an {\it extended action}, which is assumed to satisfy the {\it Batalin-Vilkovisky master equation}. BRST transformations, which include gauge transformations, are generated by the extended action, governed by BRST symmetry. Once established the master equation, the main objective is the determination of a {\it proper solution}, which is distinguished from other extended actions by suitable boundary conditions connecting it with the original action, previous to incrementation of the field content. The next goal is gauge fixing, which is nontrivially performed through the definition of a fermionic functional aimed at the elimination of the whole set of antifields. The idea is to kill two birds with one stone by getting rid of antiflields and, collaterally, fix the gauge. This process ends with the emergence of a {\it quantum action}, which depends on general gauge-fixing functions. At this point, gauge invariance has been completely removed in a general framework in which sets of {\it ad hoc} gauge-fixing functions, with minimal restrictions, can be defined to establish a particular gauge configuration.\\

To put the gauge-fixing procedure that will be introduced below in perspective, and also for clarity purposes, it is convenient to present a brief discussion about the one-loop renormalization of pure YM theories in the linear gauge~\cite{FLS}. Bare quantities are related with renormalized ones as follows: $A^{(\underline{0})a}_{{\rm B}\mu}=\sqrt{Z_A}A^{(\underline{0})a}_{\mu}$, ${C}^{(\underline{0})a}_{\rm B}=\sqrt{Z_{C}}{C}^{(\underline{0})a}$, $\bar{{C}}^{(\underline{0})a}_{\rm B}=\sqrt{Z_{ C}}\bar{{C}}^{(\underline{0})a}$, and $g_{\rm B}=\sqrt{Z_g}g$, with ${C}^{(\underline{0})a}$ and $\bar{{ C}}^{(\underline{0})a}$ the ghost and antighost fields, respectively. The bare lagrangian is given by:
\begin{equation}
{\cal L}^{(\underline{0})}_{\rm QYM,B}={\cal L}^{(\underline{0})}_{\rm QYM}+{\cal L}^{(\underline{0}){\rm YM}}_{\rm c.t.}\, ,
\end{equation}
where ${\cal L}^{(\underline{0})}_{\rm QYM}$ is the renormalized quantum Lagrangian, given by
\begin{equation}
{\cal L}^{(\underline{0})}_{\rm QYM}=-\frac{1}{4}F^{(\underline{0})a}_{\mu \nu}F^{(\underline{0}) \mu \nu}_{a}-\frac{1}{2\xi}\left(\partial_\mu A^{(\underline{0})\mu}_a\right)^2
+ \bar{{C}}^{(\underline{0})a}\left(-\partial^\mu {\cal D}^{ab}_\mu\right){C}^{(\underline{0})b}\, ,
\end{equation}
while ${\cal L}^{(\underline{0}){\rm YM}}_{\rm c.t.}$ is the counterterm Lagrangian,
\begin{eqnarray}
{\cal L}^{(\underline{0}){\rm YM}}_{\rm c.t.}&=&
\frac{\delta_A}{4}\big(\partial_\mu A^{(\underline{0})a}_\nu- \partial_\nu A^{(\underline{0})a}_\nu  \big)
\big(\partial^\mu A^{(\underline{0})a\nu}- \partial^\nu A^{(\underline{0})a\nu}\big)\nonumber \\
&&+\frac{g}{2}\big[\sqrt{Z_gZ_A}(\delta_A-1)+1 \big] \big(\partial_\mu A^{(\underline{0})a}_\nu- \partial_\nu A^{(\underline{0})a}_\nu  \big)f^{abc}A^{(\underline{0})b\mu}A^{(\underline{0})c\nu}
\nonumber \\&&
+\frac{g^2}{4}[Z_g Z_A(\delta_A-1)+1]f^{abc}A^{(\underline{0})b}_\mu A^{(\underline{0})b}_\nu f^{ade}A^{(\underline{0})d\mu}A^{(\underline{0})e\nu}\, ,
\nonumber \\&&
-\delta_{C}\, \bar{{C}}^{(\underline{0})a}\partial^2 {C}^{(\underline{0})a}-\left[\sqrt{Z_gZ_A}(\delta_{}+1)-1\right]
gf^{abc}\bar{{C}}^{(\underline{0})a}\partial^\mu(A^{(\underline{0})b}_\mu {C}^{(\underline{0})c})
\label{YMR1}
\end{eqnarray}
 with $\delta_A=Z_A-1$ and $\delta_{C}=Z_{{C}}-1$. The determination of $Z_A$, at a given order, is achieved through calculation of vacuum polarization, whereas $Z_g$, and hence the renormalized coupling constant, requires the calculation of the 3- and 4-gauge-boson vertex functions. Note that this Lagrangian has five counterterms, which depend on three renormalization parameters, namely, $Z_A$, $Z_{C}$, and $Z_g$. Thus, there must be two relations among the counterterms. Such relations emerge as a consequence of BRST symmetry governing the quantum Lagrangian. In Abelian theories, such as QED, the Ward identity ensures the equality, to all orders of perturbation theory, between the fermion self-energy counterterm and the vertex-function counterterm. This in turn implies that $Z_eZ_3=1$, where $e_{\rm B}=\sqrt{Z_e}e$ and $A_{{\rm B}\mu}=\sqrt{Z_3}A_\mu$, with $e$ and $A_\mu$ the renormalized electric charge and electromagnetic field, respectively. An important goal of the present work is the introduction of gauge-fixing procedures for the gauge fields $A^{(\underline{0})a}_\mu$ and $A^{(\underline{m})a}_\mu$, such that the relation $Z_gZ_A=1$, analogue to the aforementioned Abelian property, holds, since this will considerably simplify calculations. This relation implies that $F^{(\underline{0})a}_{{\rm B}\, \mu \nu}=\sqrt{Z_A}F^{(\underline{0})a}_{\mu \nu}$, so the counterterm of this sector becomes $-\frac{\delta_A}{4}F^{(\underline{0})a}_{\mu \nu}F^{(\underline{0})\mu \nu}_a$, which is gauge invariant. With this in mind, we turn to discuss gauge-fixing procedures for the gauge zero modes $A^{(\underline{0})a}_\mu$ and the KK excited modes $A^{(\underline{k})a}_\mu$. In both cases, ${\rm SU}(N,{\cal M}^4)$ gauge symmetry is maintained at the level of the effective action $\Gamma[A^{(\underline{0})}]$, thus simplifying the determination of the beta function and the renormalized coupling constant.\\

We write the gauge-fixing sector as ${\cal L}^{\rm GF}_{\rm KK}={\cal L}^{{\rm GF}(\underline{0})}_{\rm YM}+{\cal L}^{{\rm GF}(\underline{k})}_{\rm KK}$, with
\begin{eqnarray}
{\cal L}_{\rm YM}^{{\rm GF}(\underline{0})}&=&-\frac{1}{2\xi}f^{(\underline{0})a}f^{(\underline{0})a},
\label{LGF0}
\\
{\cal L}^{{\rm GF}(\underline{k})}_{\rm KK}&=&
-\frac{1}{2\xi}\sum_{(\underline{k})}f^{(\underline{k})a}f^{(\underline{k})a}.
\label{LGFk}
\end{eqnarray}
As displayed in Eq.~(\ref{LGF0}), gauge-fixing functions $f^{(\underline{0})a}$ define the gauge-fixing Lagrangian term ${\cal L}_{\rm YM}^{{\rm GF}(\underline{0})}$,  exclusively constituted by KK zero modes, thus being meant for the specification of a gauge configuration among those defined by the symmetry group ${\rm SU}(N,{\cal M}^4)$. On the other hand, ${\cal L}^{{\rm GF}(\underline{k})}_{\rm KK}$, shown in Eq.~(\ref{LGFk}), is a gauge-fixing Lagrangian made of both zero- and excited-mode KK fields, and which is defined by gauge-fixing functions $f^{(\underline{k})a}$. The purpose of ${\cal L}^{{\rm GF}(\underline{k})}_{\rm KK}$ is to pick a gauge configuration allowed by invariance associated to nonstandard gauge transformations.\\

Symmetry with respect to nonstandard gauge transformations can be removed from the KK effective Lagrangian ${\cal L}^{\rm YM}_{\rm QKK}$ without touching the gauge group ${\rm SU}(N,{\cal M}^4)$. The trick lies in noticing that only standard gauge transformations are associated to this 4-dimensional gauge group. In this context, a set of gauge-fixing functions $f^{(\underline{k})a}$, suitably defined to transform covariantly under ${\rm SU}(N,{\cal M}^4)$, shall get the job done. So we use the following gauge-fixing functions:
\begin{eqnarray}
f^{(\underline{k})a}&=&{\cal D}^{(\underline{0})ab}_\mu A^{(\underline{k})b\mu}-\xi m_{(\underline{k})}A^{(\underline{k})a}_{\rm G},
\label{gfc1}
\end{eqnarray}
given for the first time in Ref.~\cite{GNT}. Here ${\cal D}^{(\underline{0})ab}_\mu$ is the covariant derivative of ${\rm SU}(N,{\cal M}^{4})$, in the adjoint representation. Our choice of functions $f^{(\underline{k})a}$, given in Eq.~(\ref{gfc1}), thus leaves the issue of zero-mode gauge-fixing to the lagrangian term ${\cal L}_{\rm YM}^{{\rm GF}(\underline{0})}$.\\

Implementation of the {\it background field method}~\cite{BFM1,BFM2} on zero-mode gauge fields $A^{(\underline{0})a}_\mu$ is now carried out. To this aim, gauge fields are conveniently rescaled as $gA^{(\underline{0})a}_\mu \to A^{(\underline{0})a}_\mu$, so the YM Lagrangian becomes
\begin{equation}
-\frac{1}{4} F^{(\underline{0})a}_{\mu \nu}F^{(\underline{0})a\mu \nu}\to -\frac{1}{4g^2} F^{(\underline{0})a}_{\mu \nu}F^{(\underline{0})a\mu \nu}.
\end{equation}
Moreover, this redefinition of gauge fields $A^{(\underline{0})a}_\mu$ removes coupling-constant factors from both the YM field strength and the YM covariant derivative, which are thus given as $F^{(\underline{0})a}_{\mu \nu}=\partial_\mu A^{(\underline{0})a}_\nu- \partial_\nu A^{(\underline{0})a}_\nu+f^{abc}A^{(\underline{0})b}_\mu A^{(\underline{0})b}_\nu$ and $D^{(\underline{0})}_\mu=\partial_\mu -A^{(\underline{0})a}_\mu t^a$,
respectively, with $t^a$ the ${\rm SU}(N,{\cal M}^4)$ generators in some group representation. Next we split the gauge field $A^{(\underline{0})a}_\mu$ into a {\it classical background field}, $A^{(\underline{0})a}_\mu$, and a {\it fluctuating quantum field}, $Q^{(\underline{0})a}_\mu$, as $A^{(\underline{0})a}_\mu \to A^{(\underline{0})a}_\mu+Q^{(\underline{0})a}_\mu$.
The classical field $A^{(\underline{0})a}_\mu$ is a fixed field configuration, whereas the fluctuating field $Q^{(\underline{0})a}_\mu$ is taken as an integration variable in the functional integral. Furthermore, while $A^{(\underline{0})a}_\mu$ is found to transform as a gauge field, $Q^{(\underline{0})a}_\mu$ does it as a matter field in the adjoint representation of ${\rm SU}(N,{\cal M}^4)$. Such transformation laws are consistent with covariance of the YM curvature, expressed after splitting as $F^{(\underline{0})a}_{\mu \nu}\to F^{(\underline{0})a}_{\mu \nu} +{\cal D}^{(\underline{0})ab}_\mu Q^{(\underline{0})b}_\nu-{\cal D}^{(\underline{0})ab}_\nu Q^{(\underline{0})b}_\mu+f^{abc}Q^{(\underline{0})b}_\mu Q^{(\underline{0})c}_\nu$. In this context, we implement a gauge-fixing procedure for fluctuating quantum fields $Q^{(\underline{0})a}_\mu$. We find it convenient to fix the gauge covariantly with respect to the background gauge field, for which the zero-mode gauge-fixing functions
\begin{equation}
f^{(\underline{0})a}={\cal D}^{(\underline{0})ab}_\mu Q^{(\underline{0})b\mu},
\end{equation}
to be inserted in Eq.~(\ref{LGF0}), are introduced.\\

\subsection{New-physics effects at one loop}
Since interactions of canonical dimension greater than 4 will not be considered in the one-loop calculation, for the moment we leave this type of interactions aside and rather focus on the terms of the quantum Lagrangian defined by (\ref{QL}) with canonical dimension less than or equal to 4. As usual, we decompose the bare Lagrangian into the renormalized Lagrangian and the counterterm:
\begin{equation}
\label{BLd4}
{\cal L}^{\rm YM}_{\rm BKK\mathbf{(d\leq4) }}={\cal L}^{\rm YM}_{\rm QKK\mathbf{(d\leq4) }}+ {\cal L}^{{(\underline{0})}{\rm YM}}_{\rm c.t.}\, ,
\end{equation}
where ${\cal L}^{\rm YM}_{\rm QKK\mathbf{(d\leq4) }}$ is the renormalized quantum Lagrangian, which is given by:
\begin{equation}
\label{QLR}
{\cal L}^{\rm YM}_{\rm QKK\mathbf{(d\leq4) }}={\cal L}^{{\rm YM}(\underline{0})}_{\rm KK}+{\cal L}^{{\rm YM}(\underline{k})}_{\rm KK}
+{\cal L}^{\rm GF}_{\rm KK}+{\cal L}^{\rm G}_{\rm KK}\, .
\end{equation}
The Lagrangian term ${\cal L}^{{\rm YM}(\underline{0})}_{\rm KK}$, in the right-hand side of this equation, is interpreted as the 4-dimensional YM theory, defined, as usual~\cite{PeSch}, in 4 spacetime dimensions, but modified by the implementation of the background-field splitting previously discussed. This Lagrangian term is thus given only in terms of background fields $A^{(\underline{0})a}_\mu$ and fluctuation fields $Q^{(\underline{0})a}_\mu$. The second Lagrangian term, which has been denoted as ${\cal L}^{{\rm YM}(\underline{k})}_{\rm KK}$, is constituted by couplings in which either both KK zero and excited modes participate or only KK excited-mode fields are involved [see Eqs.(\ref{ELVV}-\ref{ELSS}), (\ref{CVV0},\ref{CVVm}), (\ref{CVV00}), (\ref{CVF}-\ref{CSF0})]. An analogous separation is implemented in the ghost-antighost sector, that is, ${\cal L}^{\rm G}_{\rm KK}={\cal L}^{{\rm G}(\underline{0})}_{\rm KK}+{\cal L}^{{\rm G}(\underline{k})}_{\rm KK}$. Then, we consider KK couplings comprised by the sum of zero-mode Lagrangian terms ${\cal L}^{{\rm YM}(\underline{0})}_{\rm KK}+{\cal L}^{{\rm GF}(\underline{0})}_{\rm KK}+{\cal L}^{{\rm G}(\underline{0})}_{\rm KK}
$. Besides fields $A^{(\underline{k})a}_\mu$ and $Q^{(\underline{0})a}_\mu$, this equation includes zero-mode ghost and antighost fields. Moreover, this sum is a gauge-fixed Lagrangian with respect to the fluctuating fields $Q^{(\underline{0})a}_\mu$; nonetheless, it is invariant under background-field gauge transformations. Note that gauge-invariance of ${\cal L}^{{\rm GF}(\underline{0})}_{\rm KK}$ implies gauge-invariance of ${\cal L}^{{\rm G}(\underline{0})}_{\rm KK}$. Taking the Feynman-'t Hooft gauge, these Lagrangian terms are written as
\begin{equation}
{\cal L}^{{\rm YM}(\underline{0})}_{\rm KK}+{\cal L}^{{\rm GF}(\underline{0})}_{\rm KK}+{\cal L}^{{\rm G}(\underline{0})}_{\rm KK}
=
\frac{1}{2g^2}Q^{(\underline{0})a}_\mu \big({\cal D}^{(\underline{0})ab}_\rho{\cal D}^{(\underline{0})bc\rho}g^{\mu \nu}+2f^{abc}F^{(\underline{0})b\mu \nu} \big)Q^{(\underline{0})c}_\nu
+\bar{C}^{(\underline{0})a}\big({\cal D}^{(\underline{0})ab}_\mu{\cal D}^{(\underline{0})bc\mu} \big)C^{(\underline{0})c}+\cdots
\label{1loopcoup0}
\end{equation}
In this expression only couplings which contribute to vacuum-polarization Feynman diagrams, at one loop, have been written explicitly, whereas the presence of other couplings is indicated by the ellipsis. Note that in the above expression explicit gauge invariance is preserved.\\

From Eq.~(\ref{QLR}), we also take the sum of Lagrangian terms
\begin{eqnarray}
{\cal L}^{{\rm YM}(\underline{k})}_{\rm KK}+{\cal L}^{{\rm GF}(\underline{k})}_{\rm KK}+{\cal L}^{{\rm G}(\underline{k})}_{\rm KK}
&=&\frac{1}{2}\sum_{(\underline{k})}A^{(\underline{k})a\mu}
\Big[
g_{\mu\nu}\big( \,{\cal D}^{(\underline{0})ab}_\rho\,{\cal D}^{(\underline{0})bc\rho}+\delta^{bc}\,m_{(\underline{k})}^2 \big)
\nonumber \\
&&-\Big( 1-\frac{1}{\xi} \,\Big){\cal D}^{(\underline{0})ab}_\mu{\cal D}^{(\underline{0})bc}_\nu
+2g_{\rm s}f^{abc}F^{(\underline{0})m}_{\mu\nu}
\Big]
A^{(\underline{k})c\nu}
\nonumber \\
&&-\frac{1}{2}\sum_{(\underline{k})}A'^{(\underline{k})a}_{\bar{n}}
\Big[\,
{\cal D}^{(\underline{0})ab}_\rho\,{\cal D}^{(\underline{0})bc\rho}+\delta^{bc}\,m_{(\underline{k})}^2
\Big]
A'^{(\underline{k})c}_{\bar{n}}
\nonumber \\
&&-\frac{1}{2}\sum_{(\underline{k})}A_{G}^{(\underline{k})a}
\Big[\,
{\cal D}^{(\underline{0})ab}_\rho\,{\cal D}^{(\underline{0})bc\rho}+\delta^{bc}\,\xi m_{(\underline{k})}^2
\Big]
A^{(\underline{k})c}_{G}
\nonumber \\
&&+\sum_{(\underline{k})}\bar{C}^{(\underline{k})a}
\Big[\,
{\cal D}^{(\underline{0})ab}_\rho\,{\cal D}^{(\underline{0})bc\rho}+\delta^{bc}\,\xi m_{(\underline{k})}^2
\Big]
C^{(\underline{k})c}+\cdots,
\label{1loopcoup}
\end{eqnarray}
which, as it occurs with the zero-mode Lagrangian (\ref{1loopcoup0}), is manifestly invariant under standard gauge transformations. In this expression,
$\bar{C}^{(\underline{k})a}$ and $C^{(\underline{k})a}$ are KK-excited-mode ghost fields, which arise as part of the quantization procedure~\cite{NT,NoTo2,GNT}. We work in the Feynman-'t Hooft gauge, which was chosen in the case of the standard theory as well. A simplification introduced by this gauge consists in the elimination, from ${\cal L}_{\rm KK}^{{\rm YM}(\underline{k})}$, of the term involving the covariant-derivatives factor ${\cal D}^{(\underline{0})}_\mu{\cal D}^{(\underline{0})}_\nu$. Moreover, in this gauge, the unphysical masses of pseudo-Goldstone bosons and ghost fields are the same as those of the KK gauge and scalar fields. The ellipsis in this equation represents other couplings, which occur either with the involvement of both KK zero and excited modes or among KK-excited-mode fields only. \\

Since explicit gauge invariance is preserved in the Lagrangians given by Eqs.~(\ref{1loopcoup0}) and (\ref{1loopcoup}), the counterterm Lagrangian  ${\cal L}^{{(\underline{0})}{\rm YM}}_{\rm c.t.}$ must be gauge invariant as well, which implies the relation $Z_AZ_g=1$. This Lagrangian can be written as the sum of a gauge sector and a ghost sector. However, in this gauge-invariant approach to quantization, the ghost fields (and also the $Q^{(\underline{0})a}_\mu$ fields) do not have to be renormalized because they only appear inside loops. So Eq.(\ref{YMR1}) reduces to
\begin{equation}
\label{CTGI}
{\cal L}^{{(\underline{0})}{\rm YM}}_{\rm c.t.}=-\frac{\delta_A}{4g^2}F^{(\underline{0})a}_{\mu \nu}F^{(\underline{0})\mu \nu}_a\, .
\end{equation}

\ \\

We define the Lagrangian term ${\cal L}_\beta$ as the sum of Eqs.~(\ref{1loopcoup0}), (\ref{1loopcoup}), and (\ref{CTGI}), but with all the terms indicated by ellipses, in such expressions, as well as those of canonical dimension higher than 4, removed. We are interested in integrating out, from ${\cal L}_\beta$, all the zero-mode fluctuating quantum fields $Q^{(\underline{0})a}_\mu$, ghost fields $C^{(\underline{0})a}$ and antighost fields $\bar{C}^{(\underline{0})a}$, together with all the KK excited modes. With this in mind, the effective action $\Gamma[A^{(\underline{0})}]$ is defined, by Eq.~(\ref{Geff}), as
\begin{eqnarray}
e^{i\Gamma[A^{(\underline{0})}]}&=&
\prod_{(\underline{k})}\prod_{x,a,\mu,\bar n}
\int {\cal D}Q^{(\underline{0})a}_\mu
{\cal D}C^{(\underline{0})a}{\cal D}\bar{C}^{(\underline{0})a}
{\cal D}A^{(\underline{k})a}_\mu
{\cal D}A'^{(\underline{k})a}_{\bar{n}}
{\cal D}A^{(\underline{k})a}_G
{\cal D}C^{(\underline{k})a}{\cal D}\bar{C}^{(\underline{k})a}\exp\big\{ i{\textstyle \int} d^4x\,{\cal L}_\beta \big\}\nonumber \\
&=&\exp\Big\{i\int d^4x\Big(\,\frac{-1}{4g^2} F^{(\underline{0})a}_{\mu \nu}F^{(\underline{0})\mu \nu}_a
+{\cal L}^{{(\underline{0})}{\rm YM}}_{\rm c.t.}\Big)\Big\}({\rm Det}\,\Delta_{Q^{(\underline{0})}})^{-\frac{1}{2}}
\, ({\rm Det}\,\Delta_{C^{(\underline{0})}})^{+1}\nonumber \\
&&\times \prod_{(\underline{m})}({\rm Det}\,\Delta_{A^{(\underline{m})}})^{-\frac{1}{2}}
({\rm Det}\,\Delta_{C^{(\underline{m})}})^{+1}
\big({\rm Det}\,\Delta_{A'^{(\underline{m})}_{\bar n}}\big)^{-\frac{\bar n}{2}}
\big({\rm Det}\,\Delta_{A^{(\underline{m})}_G}\big)^{-\frac{1}{2}}.
\label{Geff}
\end{eqnarray}
Defining the effective Lagrangian ${\cal L}^{\rm eff}_\beta$ by $\Gamma[A^{(\underline{0})}]=\exp\{i\int d^4x\,{\cal L}^{\rm eff}_\beta \}$ and performing Gaussian integrals, we write down the equation
\begin{eqnarray}
\int d^4x\,{\cal L}^{\rm eff}_\beta&=&\int d^4x\Big(\,\frac{-1}{4g^2} F^{(\underline{0})a}_{\mu \nu}F^{(\underline{0})\mu \nu}_a +{\cal L}^{{(\underline{0})}{\rm YM}}_{\rm c.t.}\Big)
+\frac{i}{2}{\rm Tr}\log\big\{ g_{\mu\nu}{\textstyle \otimes}\big({\textstyle -({\cal D}^{(\underline{0})})^2}\big)+2iF^{(\underline{0})a}_{\mu\nu}{\textstyle \otimes} T^a_{\rm a} \big\}
\nonumber \\
&&-i{\rm Tr}\log\big\{{\textstyle -({\cal D}^{(\underline{0})})^2}\big\}
+\frac{i}{2}\sum_{(\underline{k})}{\rm Tr}\log\big\{ g_{\mu\nu}{\textstyle \otimes}\big({\textstyle  -({\cal D}^{(\underline{0})})^2}
-m^2_{(\underline{k})}\cdot{\bf 1}_{\rm a} \big)+2iF^{(\underline{0})a}_{\mu\nu}{\textstyle \otimes} T^{a}_{\rm a} \big\}
\nonumber \\
&&+i\left(\frac{n}{2}-1\right)\sum_{(\underline{k})}{\rm Tr}\log\big\{{\textstyle -({\cal D}^{(\underline{0})})^2}-m_{(\underline{k})}^2\cdot{\bf 1}_{\rm a} \big\}\,.
\label{trsum}
\end{eqnarray}
In this expression, the symbol ``Tr'' denotes a trace over spacetime coordinates and over internal degrees of freedom as well. We have used {\it Kronecker-product symbols} to emphasize that the arguments of the logarithms are different of each other, living in different spaces. The notation ${\bf 1}_{\rm a}$ has been used for the identity matrix of size $(N^2-1)\times(N^2-1)$ that corresponds to the adjoint representation of the gauge group ${\rm SU}(N,{\cal M}^4)$. Note that the first and third traces, respectively coming from contributing terms with zero-mode background fields $Q^{(\underline{0})a}_\mu$ in Eq.~(\ref{1loopcoup0}) and with KK-excited-mode vector fields $A^{(\underline{k})a}_\mu$ in Eq.~(\ref{1loopcoup}), include the symbol $g_{\mu\nu}$, which should not be understood as a number, but rather as a $4\times 4$ matrix associated to the 4-dimensional Lorentz group. In the same sense, $F^{(\underline{0})a}_{\mu\nu}$ is also a $4\times4$ matrix. The second trace in Eq.~(\ref{trsum}) has been generated by the zero-mode ghost-antighost terms explicitly shown in Eq.~(\ref{1loopcoup0}).
The last trace, involving the factor $\frac{n}{2}-1$, is the total contribution from KK scalars $A^{(\underline{k})a}_{\bar{n}}$, from pseudo-Goldstone bosons $A^{(\underline{k})a}_G$, and from ghost and antighost KK-excited-mode fields $C^{(\underline{k})a}$ and $\bar{C}^{(\underline{k})a}$, all summed together. Note that, in this covariant gauge-fixing procedure, the ghost-antighost contribution is minus twice the scalar contribution.\\

Consider the Fourier transform, $\tilde{A}^{(\underline{0})a}_{\mu}(p)$, of the vector field $A^{(\underline{0})a}_\mu(x)$, defined by
\begin{equation}
A^{(\underline{0})a}_\mu(x)=\int\frac{d^4p}{(2\pi)^4}e^{-ip\cdot x}\,\tilde{A}^{(\underline{0})a}_\mu(p).
\end{equation}
In terms of this Fourier transform, the one-loop correction given by
\begin{equation}
\int d^4x\Big(-\frac{1}{4g^2}{\Pi^{\rm loop}_{\rm KK}(p^2)}\,\Big) \,F^{(\underline{0})a}_{\mu\nu}F^{(\underline{0})\mu\nu}_a
=
-\frac{1}{2g^2}\int\frac{d^4p}{(2\pi)^4}\,\tilde{A}^{(\underline{0})a}_\mu(-p)\,\tilde{A}^{(\underline{0})b}_\nu(p)
(p^2g^{\mu\nu}-p^\mu p^\nu)\delta^{ab}\,\Pi^{\rm loop}_{\rm KK}(p^2)
+\cdots
\label{YMFourier}
\end{equation}
is established. The idea is to calculate the traces in the effective Lagrangian given in Eq.~(\ref{trsum}), with the purpose of expressing their sum in the form of Eq.~(\ref{YMFourier}), aiming at the identification of the $\Pi^{\rm loop}_{\rm KK}(p^2)$ loop polarization function. The one-loop contribution to the polarization tensor $\Pi^{\mu \nu\, ab}_{\rm KK\, 1L}$ is given by:
\begin{equation}
\label{TPF1}
\Pi^{\mu \nu\, ab}_{\rm KK\, 1L}(p)=i(p^2g^{\mu\nu}-p^\mu p^\nu)\delta^{ab}\, \Pi^{\rm loop}_{\rm KK}(p^2)\, .
\end{equation}

Eqs.~(\ref{KK0gt})-(\ref{TrNS}), displayed below, provide the traces defining $\int d^4x\,{\cal L}_\beta^{\rm eff}$, in Eq.~(\ref{trsum}).
\\ \\
{\it KK-zero-modes gauge trace:}
\begin{eqnarray}
&&\frac{i}{2}{\rm Tr}\log\big\{ g_{\mu\nu}{\textstyle \otimes}\big({\textstyle -({\cal D}^{(\underline{0})})^2} \big)+2iF^{(\underline{0})a}_{\mu\nu}{\textstyle \otimes} T^a_{\rm a} \big\}
=\frac{i}{2}\int\frac{d^4p}{(2\pi)^4}\,\tilde{A}^{(\underline{0})a}_\mu(-p)\tilde{A}^{(\underline{0})a}_\nu(p)
\nonumber \\
&&\times
4N
\bigg[
-\frac{1}{2}\int\frac{d^4q}{(2\pi)^4}\,\frac{(p^\mu+2q^\mu)(p^\nu+2q^\nu)}{q^2(q+p)^2}
-\int\frac{d^4q}{(2\pi)^4}\,\frac{(p^2g^{\mu\nu}-p^\mu p^\nu)}{q^2(q+p)^2}
+g^{\mu\nu}\int\frac{d^4q}{(2\pi)^4}\,\frac{1}{q^2}
\bigg]+\cdots,
\label{KK0gt}
\end{eqnarray}
\\
{\it KK-excited-modes gauge trace:}
\begin{eqnarray}
&&\frac{i}{2}{\rm Tr}\log\big\{ g_{\mu\nu}{\textstyle \otimes}\big({\textstyle -({\cal D}^{(\underline{0})})^2}-m_{(\underline{k})}^2\cdot{\bf 1}_{\rm a} \big)+2iF^{(\underline{0})a}_{\mu\nu}{\textstyle \otimes}T^a_{\rm a} \big\}=\frac{i}{2}\int\frac{d^4p}{(2\pi)^4}\,\tilde{A}^{(\underline{0})a}_\mu(-p)\tilde{A}^{(\underline{0})a}_\nu(p)\nonumber \\
&&4N\bigg[
-\frac{1}{2}\int\frac{d^4q}{(2\pi)^4}\,\frac{(p^\mu+2q^\mu)(p^\nu+2q^\nu)}{\big[q^2-m^2_{(\underline{k})}\big]\big[(q+p)^2-m^2_{(\underline{k})}\big]}
-\int\frac{d^4q}{(2\pi)^4}\,\frac{(p^2g^{\mu\nu}-p^\mu p^\nu)}{\big[ q^2-m^2_{(\underline{k})} \big]\big[ (q+p)^2-m^2_{(\underline{k})} \big]}
\nonumber \\ &&\hspace{0.5cm}
+g^{\mu\nu}\int\frac{d^4q}{(2\pi)^4}\,\frac{1}{q^2-m^2_{(\underline{k})}}
\bigg]+\cdots,
\label{KKkgt}
\end{eqnarray}
\\
{\it KK-zero-modes ghost-antighost trace:}
\begin{eqnarray}
-i{\rm Tr}\log\big\{ -({\cal D}^{(\underline{0})})^2 \big\}
&=&-i\int\frac{d^4p}{(2\pi)^4}\,\tilde{A}^{(\underline{0})a}_\mu(-p)\tilde{A}^{(\underline{0})a}_\nu(-p)\nonumber \\
&& \times N\bigg[
-\frac{1}{2}\int\frac{d^4q}{(2\pi)^4}\,\frac{(p^\mu+2q^\mu)(p^\nu+2q^\nu)}{q^2(q+p)^2}
+g^{\mu\nu}\int\frac{d^4q}{(2\pi)^4}\frac{1}{q^2}
\bigg]+\cdots,
\label{TrS}
\end{eqnarray}
\\
{\it KK-excited-modes ghost-antighost trace:}
\begin{eqnarray}
&&-i{\rm Tr}\log\big\{ -({\cal D}^{(\underline{0})})^2-m_{(\underline{k})}^2\cdot{\bf 1}_{\rm a} \big\}
=-i\int\frac{d^4p}{(2\pi)^4}\,\tilde{A}^{(\underline{0})a}_\mu(-p)\tilde{A}^{(\underline{0})a}_\nu(-p)
\nonumber \\
&&\times
N
\bigg[
-\frac{1}{2}\int\frac{d^4q}{(2\pi)^4}\,\frac{(p^\mu+2q^\mu)(p^\nu+2q^\nu)}{\big[ q^2-m^2_{(\underline{k})} \big]\big[ (q+p)^2-m^2_{(\underline{k})} \big]}
+g^{\mu\nu}\int\frac{d^4q}{(2\pi)^4}\frac{1}{q^2-m_{(\underline{k})}^2}
\bigg]+\cdots,
\label{TrNS}
\end{eqnarray}
All the terms explicitly shown in Eqs.~(\ref{KK0gt})-(\ref{TrNS}) are given in terms of loop integrals which, after being regularized in a gauge-invariant manner and then being solved through standard methods~\cite{PeSch,Feynman,BoGi,PaVe,Weinberg}, fit into the structure of Eq.~(\ref{TPF1}). Ellipses in Eqs.~(\ref{KK0gt})-(\ref{TrNS}), on the other hand, stand for terms not contributing to the polarization function  $\Pi^{\rm loop}_{\rm KK}(p^2)$, in accordance with Eq.~(\ref{TPF1}).\\

\subsection{Divergence structure}

From Eqs.~(\ref{KK0gt})-(\ref{TrNS}), the one-loop contribution to the tensor polarization can be written as follows:
\begin{equation}
\label{1LTP1}
\Pi^{\mu \nu\, ab}_{\rm KK\, 1L}(p)=\delta^{ab}\, ig^2N\int\frac{d^4q}{(2\pi)^4}\left\{\frac{T^{\mu \nu}_V+T^{\mu \nu}_S}{q^2(q+p)^2}
+\sum_{(\underline{k})}\frac{T^{\mu \nu}_V(n)+T^{\mu \nu}_S(n)}{[q^2-m^2_{(\underline{k})}][(q+p)^2-m^2_{(\underline{k})}]} \right\}\, ,
\end{equation}
where the $T^{\mu \nu}_V$ and $T^{\mu \nu}_S$ Lorentz tensors represent the zero-mode gauge-field and ghost-antighost contributions, respectively, which are given by:
\begin{subequations}
\begin{align}
T^{\mu \nu}_V&=-2(2q+p)^\mu (2q+p)^\nu+4(p+q)^2g^{\mu \nu}-4(p^2g^{\mu \nu}-p^\mu p^\nu)\, , \\
T^{\mu \nu}_S&=(2q+p)^\mu (2q+p)^\nu-2(p+q)^2g^{\mu \nu}\, .
\end{align}
\end{subequations}
On the other hand, $T^{\mu \nu}_V(n)$ represents the gauge KK-modes $A^{(\underline{k})a}_\mu$ contribution, while $T^{\mu \nu}_S(n)$ brings together the contributions of the corresponding ghost-antighost and pseudo Goldstone-boson associated with $A^{(\underline{k})a}_\mu$, as well as those contributions from physical scalars. They are given by:
\begin{subequations}
\begin{align}
T^{\mu \nu}_V(n)&=-2(2q+p)^\mu (2q+p)^\nu+4\left[(p+q)^2-m^2_{(\underline{k})}\right]g^{\mu \nu}-4(p^2g^{\mu \nu}-p^\mu p^\nu)\, , \\
T^{\mu \nu}_S(n)&=\left(\frac{n}{2}-1\right)\left\{-(2q+p)^\mu (2q+p)^\nu+\left[2(p+q)^2-m^2_{(\underline{k})}\right]g^{\mu \nu}\right\}\, .
\end{align}
\end{subequations}

So far, we have been talking about a quantum field theory that is unusual in the sense that it comprises an infinite number of fields, while no mention to consequences that this could have on radiative corrections has been made. As already commented in the Introduction, the presence of such an infinite number of fields can become a difficult problem to handle because eventually divergences different from the usual ones may arise from the presence of discrete and continuous sums $\sum_{(\underline{k})}\int d^4k$ that emerge from excited-modes contributions. If the number of KK excitations were finite, no matter how large, we would be in a conventional scenario of calculating the one-loop contribution of a large, but finite, number of particles to a given loop amplitude.  However, in our case where the number of KK fields is infinite, the discrete sums nested in the $\sum_{(\underline{k})}$ symbol [see Eqs.(\ref{SD}) and (\ref{A2})] may or may not converge. The presence of both discrete and continuous infinite sums is strongly linked to the existence of two different spaces, namely the usual 4-dimensional spacetime manifold and the compact $n$-dimensional manifold, the presence of the infinite number of KK fields having to do precisely with this new manifold. As it is emphasized in the Introduction and commented by first time in Ref.~\cite{OPPRD}, these types of divergences can be treated as genuine ultraviolet divergences, which allow us to handle them by renormalization in the context of an effective theory~\cite{Weinberg,Burgess,Manohar,Wudka}.\\

To shed light on what the nature of the new kind of divergences that can arise from the discrete sums nested in the symbol $\sum_{(\underline{k})}$ is, we address the problem from a different perspective. The investigation of the present paper is motivated by string theory, which is the spirit of KK theories, that is, we assume that it represents the contribution of an infinite number of KK particles. However, technically speaking, the loop amplitudes, in which an infinite number of excited KK modes circulate, may be subjected to another interpretation. Before compactification, consider some particle propagating in the $4+n$ spacetime dimensions. From this point of view, the discrete momenta $q_{\bar \mu}$ can be interpreted as the components of the total momentum of such an extra-dimensional particle along the compact manifold, that is, $q_M=q_\mu +q_{\bar \mu}$. Thus, this new type of divergences, if present, can be considered as genuine ultraviolet divergences in the sense that they correspond to very-high-energy effects or, equivalently, to very-short distance effects. To show this, let us to rewrite the expression given in (\ref{1LTP1}) in the following equivalent manner:
\begin{equation}
\label{1LTP2}
\Pi^{\mu \nu\, ab}_{\rm KK\, 1L}(p)=\delta^{ab}\, ig^2N\int\frac{d^4q}{(2\pi)^4}\left\{\frac{T^{\mu \nu}_V+T^{\mu \nu}_S}{q^2(q+p)^2}
+\sum_{(\underline{q})}\frac{T^{\mu \nu}_V(n)+T^{\mu \nu}_S(n)}{[q^2-\bar{q}^2][(q+p)^2-\bar{q}^2]} \right\}\, ,
\end{equation}
where notation has been changed as $m^2_{(\underline{k})} \to  m^2_{(\underline{q})}=\bar{q}_{\bar \mu }\bar{q}_{\bar \mu }\equiv \bar{q}^2$. This expression suggests that sums over discrete squared momenta $\bar{q}^2$ comprise very-high-energy effects for KK indices $(\underline{q})$ with very large components, just as it occurs for continuous sums over very large momenta $q$. In other words, while very large continuous momenta $q_\mu$ characterize very-short-distance effects in the infinite manifold ${\cal M}^4$, very large discrete momenta $q_{\bar \mu}$ can be associated with very-short-distance effects in the compact manifold, since $1/\sqrt{\bar{q}^2}=\sqrt{R^2/\underline{q}^2} \to 0$ as $\underline{q}\to \infty$. So, when the discrete sums diverge, we have a genuine ultraviolet divergence. Later on, we will discuss how to distinguish a type of ultraviolet divergence from the other.\\

We now proceed to regularize the expression given by Eq.(\ref{1LTP2}). As emphasized in the Introduction, we will address the one-loop impact of  the  effective theory not from the cutoff approach, but using the dimensional regularization scheme~\cite{BoGi,TVreg}. In the case under consideration, we must also deal with the new type of divergences that can arise from discrete sums. Our approach consists in regularizing simultaneously both types of divergent quantities, because, as previously argued, they have the same origin in the sense that the sums in consideration involve the magnitudes $q_\mu$ and $\bar{q}_{\bar \mu}$, which are linked to the usual and compact manifolds through Fourier transform ($q_\mu$) and Fourier series ($\bar{q}_{\bar \mu}$). The main idea behind this is to express the regularized $\sum_{(\underline{k})}$ sums in terms of Epstein zeta functions~\cite{E1,E2,E3} defined in the complex plane through analytic continuation. The inhomogeneous $l$-dimensional Epstein function is defined as
\begin{equation}
\label{EF}
E^{c^2}_l(s)=\sum^\infty_{(k_1,\cdots,k_l)=1}\frac{1}{\left(k^2_1+\cdots+k^2_l+c^2\right)^s}\, .
\end{equation}
A special case corresponds to $c=0$, which leads to the homogeneous Epstein function. \\

As usual, we promote the ordinary 4-dimensional spacetime to $D$ dimensions. After that and once Feynman parametrization has been implemented,  Eq.~(\ref{1LTP2}) becomes
\begin{equation}
\label{1LTP3}
\Pi^{\mu \nu\, ab}_{\rm KK\, 1L}(p)=\delta^{ab}\frac{g^2N}{(4\pi^2)}\int^1_0dx \frac{(4\pi \hat{\mu}^2)^{2-\frac{D}{2}}}{i\pi^{\frac{D}{2}}}\int d^Dq\left\{\frac{T^{\mu \nu}_{(\underline{0})}}{\left(q^2-\Delta^2_{(\underline{0})}\right)^2}+\sum_{(\underline{k})}\frac{T^{\mu \nu}_{(\underline{k})}}{\left(q^2-\Delta^2_{(\underline{k})}\right)^2}\right\}\, ,
\end{equation}
where $\hat{\mu}$ is the scale associated with dimensional regularization, $\epsilon=4-D$, $\Delta^2_{(\underline{0})}=-x(1-x)p^2$, and $\Delta^2_{(\underline{k})}=m^2_{(\underline{k})}+\Delta^2_{(\underline{0})}$. In addition,
\begin{subequations}
\begin{align}
T^{\mu \nu}_{(\underline{0})}&=-\left[2\left(1-\frac{2}{D}\right)q^2+2(1-x)^2p^2\right]g^{\mu \nu}+(1-2x)^2p^\mu p^\nu+4(p^2g^{\mu \nu}-p^\mu p^\nu)\\
T^{\mu \nu}_{(\underline{0})}&=2\left(1+\frac{D}{2}\right)\left[\left(\frac{2}{D}-1\right)q^2-(1-x)^2p^2+m^2_{(\underline{k})}\right]g^{\mu \nu}+\left(1+\frac{n}{2}\right)(1-2x)^2p^\mu p^\nu +4(p^2g^{\mu \nu}-p^\mu p^\nu)\, .
\end{align}
\end{subequations}
Once the integrals on the momentum space have been performed, the one-loop polarization function can be expressed as follows:
\begin{equation}
\label{1LTP4}
\Pi^{\rm loop}_{\rm KK}(p^2)=\frac{g^2N}{(4\pi )^2}\int^1_0dx\left\{f(x)\Gamma\left(\frac{\epsilon}{2}\right)\left(\frac{\Delta^2_{(\underline{0})}}{4\pi \hat{\mu}^2}\right)^{-\frac{\epsilon}{2}}+
\left[f(x)-\frac{n}{2}g(x)\right]\sum_{(\underline{k})}\Gamma\left(\frac{\epsilon}{2}\right)\left(\frac{\Delta^2_{(\underline{k})}}{4\pi \hat{\mu}^2}\right)^{-\frac{\epsilon}{2}} \right\}\, ,
\end{equation}
where $f(x)=4-g(x)$, with $g(x)=(1-2x)^2$. Note that the KK-modes contribution contains, besides the gauge contribution characterized by the $f(x)$ function, the contribution of $n$ scalar matter fields, which correspond to the $n-1$ physical scalars $A^{(\underline{k})a}_{\bar n}$ and the pseudo Goldstone boson $A^{(\underline{k})a}_G$ associated with the $A^{(\underline{k})a}_\mu$ gauge boson. Note now that
\begin{eqnarray}
\sum_{(\underline{k})}\left(\frac{\Delta^2_{(\underline{k})}}{4\pi \hat{\mu}^2}\right)^{-\frac{\epsilon}{2}}&=&\left(\frac{R^{-2}}{4\pi\hat{\mu}^2}\right)^{-\frac{\epsilon}{2}}
\sum_{(\underline{k})}\left(\underline{k}^2+c^2\right)^{-\frac{\epsilon}{2}}\nonumber \\
&=&\left(\frac{R^{-2}}{4\pi\hat{\mu}^2}\right)^{-\frac{\epsilon}{2}}\sum^n_{l=1}\left(\begin{array}{ccc}
n \\
l
\end{array}\right)E^{c^2}_l\left(\frac{\epsilon}{2}\right)\, ,
\end{eqnarray}
where in the last step, Eqs. (\ref{A2}) and (\ref{EF}) have been used. In addition, $c^2=\frac{\Delta^2_{(\underline{0})}}{R^{-2}}$, where, for the sake of simplicity, the same radii for the $n$ orbifolds $S^1/Z_2$, $R\equiv R_1=\cdots=R_n$, have been assumed. Using this expression, the polarization function (\ref{1LTP4}) becomes
\begin{eqnarray}
\label{1LTP5}
\Pi^{\rm loop}_{\rm KK}(p^2)&=&\frac{g^2N}{(4\pi )^2}\int^1_0dx\Bigg\{f(x)\Gamma\left(\frac{\epsilon}{2}\right)\left(\frac{\Delta^2_{(\underline{0})}}{4\pi \hat{\mu}^2}\right)^{-\frac{\epsilon}{2}}\nonumber \\
&&+\left[f(x)-\frac{n}{2}g(x)\right]\left(\frac{R^{-2}}{4\pi\hat{\mu}^2}\right)^{-\frac{\epsilon}{2}}\sum^n_{l=1}\left(\begin{array}{ccc}
n \\
l
\end{array}\right)\Gamma\left(\frac{\epsilon}{2}\right)E^{c^2}_l\left(\frac{\epsilon}{2}\right) \Bigg\}\, .
\end{eqnarray}
Since the $l$-dimensional Epstein function $E^{c^2}_l(s)$ has poles at $s=\frac{l}{2}, \frac{l-1}{2},\cdots, -\frac{1}{2}, -\frac{3}{2}, \cdots$, except zero~\cite{K}, it is clear that  $E^{c^2}_l\left(\frac{\epsilon}{2}\right)$ converges for $\epsilon \to 0$. However, the way this happens is subtle, which can be appreciated more clearly by expressing the Epstein function in terms of the Riemann zeta function, whose properties are well known in the literature. An important result~\cite{E3} consists in expressing the $l$-dimensional Epstein function in terms of the one-dimensional Epstein function as
\begin{equation}
\label{EFR}
E^{c^2}_l(s)=\frac{(-1)^{l-1}}{2^{l-1}}\sum^{l-1}_{p=0}\left(\begin{array}{ccc}
l-1 \\
p
\end{array}\right)(-1)^p \pi^{\frac{p}{2}}\frac{\Gamma\left(s-\frac{p}{2}\right)}{\Gamma(s)}E^{c^2}_1\left(s-\frac{p}{2}\right)\, .
\end{equation}
The reduction to the Riemann zeta function is achieved through the following power series in $c^2$~\cite{ER,OPPRD}:
\begin{equation}
\label{EF1}
E^{c^2}_1(s)=\sum^\infty_{k=0}\frac{(-1)^k}{k!}\frac{\Gamma(k+s)}{\Gamma(s)}\zeta(2k+2s)c^{2k}\, ,
\end{equation}
where the Riemman function is defined by
\begin{equation}
\label{RF}
\zeta(s)=\sum^\infty_{n=1}\frac{1}{n^s}\, ,
\end{equation}
which has a simple pole at $s=1$. We can see, from Eqs. (\ref{EFR}) and (\ref{EF1}), that finite terms which are the ratio of two quantities that diverge when $s$ tends to zero, as $ \frac{\zeta(1+s)}{\Gamma(s)}$, arise for $l>1$. This fact has nontrivial consequences when a product of the form $\Gamma(s)E^{c^2}_l(s)$, as the one appearing in the polarization function (\ref{1LTP5}), is considered. Using the above results, we write

\begin{equation}
\label{GR}
\sum^{n}_{l=1}\left(\begin{array}{ccc}
n \\
l
\end{array}\right)\Gamma\left(\frac{\epsilon}{2}\right)E^{c^2}_l\left(\frac{\epsilon}{2}\right)=\frac{1}{2^{n-1}}\sum^n_{r=1}\sum^r_{l=1}\left(\begin{array}{ccc}
n \\
l-1
\end{array}\right)\pi^{\frac{n-r}{2}}\sum^\infty_{k=0}\frac{(-1)^k}{k!}\Gamma\left(\frac{2k+r-n+\epsilon}{2}\right)\zeta(2k+r-n+\epsilon)c^{2k}\, .
\end{equation}
Note that this power series in $c^2$ is actually a power series in the external momentum $p^2$. A careful analysis of this expression leads us to conclude that there are two types of divergences for $\epsilon \to 0$. One type of divergences, whose main feature is to be independent of both the external momenta and the compactification scale, can easily be identified from the $k=0$ term in (\ref{GR}). This type of divergences, which do not depend on the external momentum, are usual ultraviolet divergences that emerge from short-distance effects in the standard manifold ${\cal M}^4$. Another type of ultraviolet divergences, different to the usual ones in the sense that they emerge as coefficients of the external momentum or, more precisely, from the ratio $\frac{p^2}{R^{-2}}$, occurs for terms with $k\neq 0$ in the series (\ref{GR}). This can happen for an even integer $2k+r-n\leq 0$ or for $2k+r-n=1$, which correspond to poles, when $\epsilon \to 0$, of the gamma and Riemann functions, respectively. Note that the products of the form $\Gamma\left(\frac{2k+r-n+2}{2}\right)\zeta(2k+r-n+2)$ converge when $2k+r-n+2=-2,-4,-\cdots$, since the negative even integers correspond to the so-called trivial zeros of the Riemann zeta function. These types of divergences, which arise from non-convergence of products $\Gamma\left(\frac{\epsilon}{2}\right)E^{c^2}_l\left(\frac{\epsilon}{2}\right)$, can also be considered as genuine ultraviolet divergences because, as already commented, they correspond to effects of large discrete momentum or, equivalently, to short distances effects in the compact manifold. For the analysis that follows, it is convenient to rewrite the above expression so that it explicitly shows those terms that are divergent. Once this rearrangement is implemented, we get
\begin{eqnarray}
\label{GR1}
\sum^{n}_{l=1}\left(\begin{array}{ccc}
n \\
l
\end{array}\right)\Gamma\left(\frac{\epsilon}{2}\right)E^{c^2}_l\left(\frac{\epsilon}{2}\right)&=&g_{(0)}(n)\,
\Gamma\left(\frac{\epsilon}{2}\right)\zeta(\epsilon)+F_{(0)}(n)+\sum^{[\frac{n}{2}]}_{k=1}\Big[f_{(k)}(n)\frac{1}{\sqrt{\pi}}
\Gamma\left(\frac{1+\epsilon}{2}\right)\zeta(1+\epsilon)\nonumber \\
&& +g_{(k)}(n)\Gamma\left(\frac{\epsilon}{2}\right)\zeta(\epsilon)+F_{(k)}(n)  \Big]c^{2k}+F(n,c^2)\, ,
\end{eqnarray}
where
\begin{subequations}
\begin{align}
g_{(0)}(n)&=2\left(1-\frac{1}{2^n}\right)\, , \\
F_{(0)}(n)&=\frac{1}{2^{n-1}}\sum^{n-1}_{r=1}\sum^{r}_{l=1}\left(\begin{array}{ccc}
n \\
l-1
\end{array}\right)\pi^{\frac{n-r}{2}}\Gamma\left(\frac{r-n}{2}\right)\zeta(r-n)\, , \, \, n>1\\
F_{(k)}(n)&=\frac{1}{2^{n-1}}\left(\sum^n_{r\neq n-(2k-1),n-2k} \right)\sum^r_{l=1}\left(\begin{array}{ccc}
n \\
l-1
\end{array}\right)\pi^{\frac{r-n}{2}}\Gamma\left(\frac{2k+r-n}{2}\right)\zeta(2k+r-n)\, , \, \, n>1\\
\label{FFu}
F(n,c^2)&=\frac{1}{2^{n-1}}\sum^n_{r=1}\sum^{r}_{l=1}\left(\begin{array}{ccc}
n \\
l-1
\end{array}\right)\pi^{\frac{n-r}{2}}\sum^{\infty}_{k=[\frac{n}{2}]+1}\frac{(-1)^k}{k!}\Gamma\left(\frac{2k+r-n}{2}\right)\zeta(2k+r-n)c^{2k}\, .
\end{align}
\end{subequations}
In the above expressions, the symbol $[\frac{n}{2}]$ means the {\it floor of} $\frac{n}{2}$, that is, the largest integer less than or equal to $\frac{n}{2}$. On the other hand, the functions $f_{(k)}(n)$ and $g_{(k)}(n)$ are given by:
\begin{subequations}
\begin{align}
f_{(k)}(n)=& \frac{1}{2^{n-1}}\sum^{n-2k+1}_{l=1}\left(\begin{array}{ccc}
n \\
l-1
\end{array}\right)\frac{(-1)^k}{k!}\pi^{\frac{2k-1}{2}}\, , \\
g_{(k)}(n)=& \frac{1}{2^{n-1}}\sum^{n-2k}_{l=1}\left(\begin{array}{ccc}
n \\
l-1
\end{array}\right)\frac{(-1)^k}{k!}\pi^{k}\, .
\end{align}
\end{subequations}
These functions have the following properties:
\begin{subequations}
\begin{align}
&f_{(k)}(1)=\cdots=f_{(k)}(2k-1)=0\, , \\
&g_{(k)}(1)=\cdots=g_{(k)}(2k)=0\, .
\end{align}
\end{subequations}
These relations imply that in the case of only one extra dimension no divergences associated with the power series in $p^2$ emerge, since in this case $f_{(k)}(1)=0$ and $g_{(k)}(1)=0$ for all $k=1,2,\cdots$. In the case $n=2$, $g_{(k)}(2)=0$ for all $k$, but $f_{(1)}(2)\neq 0$ and $f_{(k)}(2)= 0$ for $k=2,3,\cdots$. If $n=3$, besides $f_{(1)}(3)\neq 0$, we have $g_{(1)}(3)\neq 0$, but $f_{(k)}(3)=0$ and $g_{(k)}(3)=0$ for all $k=2,3, \cdots$. Thus, for $n=2$ and $n=3$ divergences arise only in the first term ($k=1$) of the power series in $p^2$. However, for the cases $n=4$ and  $n=5$, divergences arise in the first $(k=1)$ and second $(k=2)$ terms of the power series in $p^2$; the cases $n=6$ and $n=7$ lead to divergences in the first $(k=1)$, second $(k=2)$, and third $(k=3)$ terms of the power series in $p^2$; and so on. As we will see below, the (\ref{GR1}) decomposition of the multidimensional Epstein functions will play and central role in our analysis.\\

As already commented, the case $n=1$ is the only one which does not have ultraviolet divergences arising from short-distance effects in the compact manifold. In this case, the expression (\ref{GR1}) becomes
\begin{equation}
\label{S5D}
\Gamma\left(\frac{\epsilon}{2}\right)E^{c^2}_1\left(\frac{\epsilon}{2}\right)=\Gamma\left(\frac{\epsilon}{2}\right)\zeta(\epsilon)+F(1,c^2)\, ,
\end{equation}
where the $F(1,c^2)$ function is free of ultraviolet divergencies and is given by
\begin{eqnarray}
\label{S5D1}
F(1,c^2)&=&\sum^\infty_{s=1}\frac{(-1)^s}{s}\zeta(2s)c^{2s}\nonumber \\
&=&-\sum^\infty_{k=1}\log\left(1+\frac{c^2}{k^2}\right)\nonumber \\
&=&\log\left(|\Gamma(1+ic)|^2\right)\, ,
\end{eqnarray}
where in the last step we used the the software {\it Mathematica}, by Wolfram~\cite{Math}, to perform the infinite sum.


\section{Vacuum polarization and beta function}
\label{AF}
Asymptotic freedom is a physical phenomenon in which an interacting theory becomes asymptotically noninteracting because of some coupling constant that decreases as shorter distances are explored. This aspect, known to characterize non-Abelian gauge theories~\cite{GrWi,Politzer,tHooft}
and which is central for understanding the strong interaction, is studied through renormalization-group techniques.
It turns out that beta functions corresponding to non-Abelian gauge theories that include a sufficiently-small number of fermions are negative, which thus yield the asymptotic-freedom behavior. The purpose of this section is to investigate one-loop impact of universal extra dimensions on the beta function.\\

The case $n = 1$ involves features that make it worthy of special attention, so we will divide our analysis into two scenarios, one with only one extra dimension and the other with an arbitrary number $n\geq 2$ of extra dimensions.\\

\subsection{Case $n=1$}
As already discussed, in the case of one extra dimension no ultraviolet divergences from short-distance effects in the compact manifold arise. In other words, the coefficients of $p^{2k}$ (or $c^{2k}$) in (\ref{GR1}) are all finite. As we will see, this theory has much in common with conventional field theories. An important aspect is that any vertex function of canonical dimension higher than four is free of ultraviolet divergences, which has been explicitly verified in many phenomenological studies~\cite{PHS1,PHS2,PHS3,PHS4,PHS5}. This means that, if only Green's functions with zero-mode fields as external legs are considered and not interactions of canonical dimension higher than four are inserted in loop diagrams, the theory given by the quantum Lagrangian (\ref{QLR}) is self-contained, with a counterterm given by Eq. (\ref{CTGI}) if a gauge-invariant quantization procedure has been implemented, as in the case at hand.\\

From Eqs. (\ref{CTGI}) and (\ref{1LTP5}), the renormalized polarization function is given by
\begin{eqnarray}
\label{RPF5D}
\Pi^{\rm 5D}_{\rm KK}(p^2)&=&\frac{g^2N}{(4\pi )^2}\int^1_0dx\bigg\{f(x)\Gamma\left(\frac{\epsilon}{2}\right)\left(\frac{\Delta^2_{(\underline{0})}}{4\pi \hat{\mu}^2}\right)^{-\frac{\epsilon}{2}}
+\left[f(x)-\frac{1}{2}g(x)\right]\left(\frac{R^{-2}}{4\pi\hat{\mu}^2}\right)^{-\frac{\epsilon}{2}}
\Gamma\left(\frac{\epsilon}{2}\right)E^{c^2}_1\left(\frac{\epsilon}{2}\right) \bigg\}-\delta_A\, ,
\end{eqnarray}
with $\Gamma\left(\frac{\epsilon}{2}\right)E^{c^2}_1\left(\frac{\epsilon}{2}\right)$ given by Eqs.~(\ref{S5D}) and (\ref{S5D1}). In this expression, $\delta_A$ is the contribution of the gauge-invariant counterterm (\ref{CTGI}). Note that, in this case, the only contribution from matter fields comes from pseudo-Goldstone bosons $A^{(\underline{k})a}_{\rm G}$, associated to the gauge fields $A^{(\underline{k})a}_\mu$.\\

In usual ${\rm YM}$ theories, the beta function is calculated by using a mass-independent renormalization scheme, such as ${\rm MS}$ or ${\rm \overline{MS}}$. In pure  ${\rm YM}$ theories,  without matter fields, the beta function, besides gauge independent, must be scale independent. However, in our case, the mass spectrum of the theory comprises a wide range of energies, thus suggesting that usage of a mass-dependent scheme that allows us to study the sensitivity of beta function to new-physics effects is suitable. To contrast differences, we will calculate the beta function in both types of schemes.\\

\noindent \textit{ Mass-independent scheme}. We use the ${\rm MS}$ scheme, in which the counterterm is determined by the pole of the divergence. From Eqs. (\ref{RPF5D}) and (\ref{S5D}), we have
\begin{equation}
\label{CTMS5}
\delta_A=\frac{g^2N}{(4\pi)^2}\left[\frac{11}{3}+\left(\frac{11}{3}-\frac{1}{6}\right)\zeta(0)\right]\left(\frac{2}{\epsilon}\right)\, ,
\end{equation}
where the contributions from the gauge field $A^{(\underline{k})a}_\mu$ and its pseudo-Goldstone boson $A^{(\underline{k})a}_{\rm G}$ have been explicitly displayed. This aims at emphasizing that the longitudinal components of the gauge fields behave as matter fields, since their contributions have opposite signs to the contributions from the transverse polarization states. It is important to note that $\zeta(0)=\sum^\infty_{k=1}=-1/2$ quantifies the usual ultraviolet divergences induced by the infinite number of KK excited modes.\\

Using that, in this scheme, the one-loop beta function can be calculated from the renormalization factor $Z_A$ in the form: $\beta(g)=-\frac{1}{2}g^2 \frac{\partial Z_A}{\partial g}=-\frac{1}{2}g^2\frac{\partial \delta_A}{\partial g}$, we then find that
\begin{eqnarray}
\label{BMS}
\beta_{\rm 5D}(g)&=&\beta(g)\left[1+\left(\frac{21}{22}\right)\zeta(0)\right]\nonumber \\
&=&\left(\frac{23}{44}\right)\beta(g)\, ,
\end{eqnarray}
where $\beta(g)=-\frac{g^3}{(4\pi)^2}\left(\frac{11N}{3}\right)$ is the usual one-loop beta function. From this result, we can appreciate that KK excited modes globally contribute to the beta function as matter fields. Operatively, this is caused by the negative value of the Riemann function $\zeta(0)$.\\

In this scheme, the counterterm is given by
\begin{eqnarray}
\label{PFMS}
\Pi^{\rm 5D}_{\rm KK}(p^2)&=&-\frac{g^2N}{(4\pi)^2}\int^1_0dx \bigg\{f(x)\log\left(\frac{\Delta^2_{(0)}}{4\pi e^{-\gamma} \hat{\mu}^2}\right)\nonumber \\
&&+\left[f(x)-\frac{1}{2}g(x)\right]\left[\frac{1}{2}\log\left(\frac{R^{-2}}{16\pi^3 e^{-\gamma}\hat{\mu}^2}\right)+
\log\left(\left|\Gamma(1+ic)\right|^2\right)\right]\bigg\}\, .
\end{eqnarray}
Note that new-physics effects in both the beta function and the renormalized polarization function are non-decoupling, which is a disconcerting result, since effects of heavy physics must decouple. This suggests that a mass-dependent renormalization scheme must be used.\\

\noindent \textit{Mass-dependent scheme}. The result obtained above for the beta function in a mass-independent scheme, which does not depend on the compactification scale $R^{-1}$, is surprising because, on physical grounds, one would expect new-physics effects to decouple in the limit of a very large compactification scale. While simple to implement, mass-independent schemes have the serious disadvantage that heavy particles do not decouple at energies lower their masses, in accordance with the Appelquist-Carazzone decoupling theorem~\cite{ACDT}. Here, we recalculate the beta function by using a mass-dependent scheme. To determine the counterterm $\delta_A$ we use the renormalization condition
\begin{equation}
\Pi^{\rm 5D}_{\rm KK}(p^2=-\mu^2)=0\, ,
\end{equation}
where $\mu$ is the scale of the kinematical point. This condition leads to the following counterterm:
\begin{equation}
\label{CT}
\delta_A=\frac{g^2N}{(4\pi )^2}\int^1_0dx\left\{f(x)\Gamma\left(\frac{\epsilon}{2}\right)\left(\frac{\bar{\Delta}^2_{(0)}}{4\pi \hat{\mu}^2}\right)^{-\frac{\epsilon}{2}}
+\left[f(x)-\frac{1}{2}g(x)\right]\left(\frac{R^{-2}}{4\pi\hat{\mu}^2}\right)^{-\frac{\epsilon}{2}}
\Gamma\left(\frac{\epsilon}{2}\right)E^{\bar{c}^2}_1\left(\frac{\epsilon}{2}\right) \right\}\, ,
\end{equation}
where $\bar{\Delta}^2_{(0)}=x(1-x)\mu^2$ and $\bar{c}^2=\frac{\bar{\Delta}^2_{(0)}}{R^{-2}}$. The one-loop beta function is given by $\beta_{\rm 5D}(g)=g\mu^2 \frac{\partial\delta_A}{\partial \mu^2}$, so taking into account that $\mu^2 \frac{\partial}{\partial \mu^2}\left(\frac{\bar{\Delta}^2_{(0)}}{4\pi \hat{\mu}^2}\right)^{-\frac{\epsilon}{2}}=-\frac{\epsilon}{2}$ and that $\mu^2 \frac{\partial E^{\bar{c}^2}_1(\frac{\epsilon}{2})}{\partial \mu^2}=-\frac{\epsilon}{2}\bar{c}^2E^{\bar{c}^2}_1\left(1+\frac{\epsilon}{2}\right)$, we then find
\begin{eqnarray}
\beta_{\rm 5D}(g)&=&\beta(g)-\frac{g^3N}{(4\pi)^2}\int^1_0 dx\left[f(x)-\frac{1}{2}g(x)\right]\bar{c}^2E^{\bar{c}^2}_1(1)\nonumber \\
&=&\beta(g)-\frac{g^3N}{(4\pi)^2}\int^1_0 dx\left[f(x)-\frac{1}{2}g(x)\right]\sum^\infty_{k=1}\frac{x(1-x)\mu^2}{m^2_{(k)}+x(1-x)\mu^2}\, .
\end{eqnarray}
First all, note that the usual one-loop beta function of a pure ${\rm YM}$ theory coincides in both mass-independent and mass-dependent schemes, which implies that it is scale independent. As far as the KK excited-modes contribution is concerned, it is given by a growing positive function on $\mu^2/R^{-2}$, which can be explicitly calculated~\cite{Math}:
\begin{equation}
\sum^\infty_{k=1}\frac{\bar{c}^2}{k^2+\bar{c}^2}=\frac{1}{2}\left[\pi \bar{c}\coth(\pi\bar{c})-1\right]\, .
\end{equation}
Note that this function vanishes in the limit as $\bar{c}\to 0$, which is expected from the decoupling theorem~\cite{ACDT}. Also, note that it diverges for large $\bar{c}$, though keep in mind that such a scenario is beyond the range of validity of our effective theory, which is valid for energies $\mu$ lower than the compactification scale $R^{-1}$. Since $\bar{c}^2\ll 1$, this function can be expressed as a power series in $\bar{c}^2$, which behaves as  $\bar{c}^2E^{\bar{c}^2}_1(1)=\zeta(2)\bar{c}^2-\zeta(4)\bar{c}^4+\cdots$. At first order in $\bar{c}^2$, the beta function can be written as
\begin{equation}
\beta_{\rm 5D}(g)=\beta(g)\left[1+\left(\frac{37}{220}\right)\zeta(2)\left(\frac{\mu^2}{R^{-2}}\right)+\cdots\right]\, .
\end{equation}
It is important to note that both the zero- and excited-modes contributions have the same sign. Furthermore, the correction is of the order of $(0.276648)\left(\frac{\mu^2}{R^{-2}}\right)$.\\

From Eqs. (\ref{RPF5D}) and (\ref{CT}), the renormalized polarization function can be written in the $\mu$-scheme as follows:
\begin{eqnarray}
\label{RPF5D1}
\Pi^{\rm 5D}_{\rm KK}(p^2)&=&\frac{g^2N}{(4\pi )^2}\int^1_0dx\left\{f(x)\log\left(\frac{\bar{\Delta}^2_{(0)}}{\Delta^2_{(0)}}\right)+\left[f(x)-\frac{1}{2}g(x)\right]\sum^\infty_{k=1}
\log\left(\frac{m^2_{(k)}+\bar{\Delta}^2_{(0)}}{m^2_{(k)}+\Delta^2_{(0)}}\right)
\right\}\nonumber \\
&=&\frac{g^2N}{(4\pi )^2}\int^1_0dx\left\{f(x)\log\left(\frac{\bar{\Delta}^2_{(0)}}{\Delta^2_{(0)}}\right)+\left[f(x)-\frac{1}{2}g(x)\right]
\log\left(\left|\frac{\Gamma\left(1+i\Delta^2_{(0)}\right)}{\Gamma\left(1+i\bar{\Delta}^2_{(0)}\right)}\right|^2\right)
\right\}\, .
\end{eqnarray}
The last term of this expression clearly shows the decoupling nature of new-physics effects, since this term vanishes in the limit of a very large compctification scale $R^{-1}$. This result must be compared with that obtained from the MS scheme. \\

\subsection{Case $n\geq 2$}
Previously, we have shown that in a ${\rm YM}$ theory with only one extra dimension the usual counterterm $\delta_A$ is enough to remove divergences if no loop insertions of canonical dimension higher than four are introduced. Also, we have emphasized that for a number of extra dimensions equal to or higher than two, additional divergences arise as coefficients of powers of external momentum. We have argued that since this type of divergences are associated with discrete sums, they must be attributed to short-distance effects in the compact manifold. However, these divergences cannot be removed by the usual counterterm $\delta_A$, which means that additional interactions introducing the required counterterms must be considered. Because such divergences multiply powers of external momentum, the new type of interactions must be of canonical dimension higher than four. Of course, this type of interactions are not renormalizable in the power-counting sense, but they are renormalizable in a wider sense, since our effective Lagrangian includes all the interactions allowed by symmetries, so there is a counterterm available to cancel any divergence~\cite{Weinberg,Burgess,Manohar,Wudka}. Due to gauge invariance, the building blocks of such interactions must be the ${\rm YM}$ curvature and its covariant derivatives. With this in mind and taking into account the result given by Eq.~(\ref{GR1}), the required bare Lagrangian is of the form:
\begin{equation}
{\cal L}^{{\rm YM}(4+n)}_{\rm B\,KK}= {\cal L}^{\rm YM}_{\rm QKK\mathbf{(d\leq4) }}+{\cal L}^{{\rm YM} (\underline{0})}_{{\rm KK}(\mathbf{d>4})}+
{\cal L}^{{\rm YM} (\underline{0})}_{\rm c.t.}+{\cal L}^{{\rm YM} (\underline{0})}_{{\rm c.t.}(\mathbf{d>4})}\, ,
\end{equation}
where  ${\cal L}^{\rm YM}_{\rm QKK\mathbf{(d\leq4) }}$ and ${\cal L}^{{\rm YM} (\underline{0})}_{\rm c.t.}$ are given by Eqs.~(\ref{QLR}) and (\ref{CTGI}), respectively. Moreover, ${\cal L}^{{\rm YM} (\underline{0})}_{{\rm KK}(\mathbf{d>4})}$ represents interactions of canonical dimension higher than four, which induce the counterterms needed to remove the divergences generated at one loop by the Lagrangian ${\cal L}^{\rm YM}_{\rm QKK\mathbf{(d\leq4) }}$. Such a Lagrangian can be written as
\begin{equation}
\label{YMNR}
{\cal L}^{{\rm YM} (\underline{0})}_{{\rm KK}(\mathbf{d>4})}=\sum^{[\frac{n}{2}]}_{k=1}\frac{\lambda_{(k)}}{(R^{-2})^k}\left({\cal D}_{\alpha_1}\cdots{\cal D}_{\alpha_k}F_{\mu \nu}\right)^a\left({\cal D}^{\alpha_1}\cdots{\cal D}^{\alpha_k}F^{\mu \nu}\right)^a \, .
\end{equation}
Note that our gauge-invariant renormalization scheme implies that ${\cal D}_{{\rm B}\alpha_i}={\cal D}_{\alpha_i}$, since
$Z_AZ_g=1$. So, the corresponding counterterm is gauge invariant as well and is given by
\begin{equation}
\label{CTNR}
{\cal L}^{{\rm YM} (\underline{0})}_{{\rm c.t.}(\mathbf{d>4})}=\sum^{[\frac{n}{2}]}_{k=1}\frac{\delta^{(k)}_{A}}{(R^{-2})^k}\left({\cal D}_{\alpha_1}\cdots{\cal D}_{\alpha_k}F_{\mu \nu}\right)^a\left({\cal D}^{\alpha_1}\cdots{\cal D}^{\alpha_k}F^{\mu \nu}\right)^a \, ,
\end{equation}
where $\delta^{(k)}_A\equiv Z_A\lambda_{B(k)}-\lambda_{(k)}$.\\

Then, up to one-loop order, the contribution from the Lagrangian (\ref{YMNR}) to the polarization function is given by
\begin{equation}
\Pi^{(4+n){\rm D}}_{\rm KK}(p^2)=\sum^{[\frac{n}{2}]}_{k=1}\lambda_{(k)}\left(\frac{p^2}{R^{-2}}\right)^k +\Pi^{\rm loop}_{\rm KK}(p^2)-\delta_A+\sum^{[\frac{n}{2}]}_{k=1}\delta^{(k)}_A\left(\frac{p^2}{R^{-2}}\right)^k\, ,
\end{equation}
where the first term of this expression is the tree-level contribution of the Lagrangian (\ref{YMNR}), the second term represents the one-loop  contribution induced by the Lagrangian (\ref{QLR}), which is given by Eq.~(\ref{1LTP5}), the third term is the contribution of the usual counterterm, and the last term represents the contribution of the counterterms induced by interactions of dimension higher than four. To determine the $[\frac{n}{2}]+1$ counterterms, we proceed as in the case $n=1$, discussed above, by using both the MS scheme and the $\mu$ scheme. \\

\noindent \textit{Mass-independent scheme}. The counterterms are determined by the coefficients of the poles of the divergences:
\begin{subequations}
\begin{align}
\delta_A&=\frac{g^2N}{(4\pi)^2}\left[\frac{11}{3}+
\left(\frac{11}{3}-\frac{n}{6}\right)\zeta(0)\right]\left(\frac{2}{\epsilon}\right)\, , \\
\delta^{(k)}_A&=-\frac{g^2N}{(4\pi)^2}\sum^{[\frac{n}{2}]}_{k=1}(-1)^k\left[f_{(k)}(n)-g_{(k)}(n)\right]\int^1_0dx\left[f(x)-\frac{n}{2}g(x)\right]
[x(1-x)]^k\, .
\end{align}
\end{subequations}
Note that the counterterm $\delta_A$ in this expression is identical to that of Eq.~(\ref{CTMS5}), except for the value of $n$. Consequently, the beta function is given by $\beta_{(4+n){\rm D}}=\beta \left(\frac{22+n}{44}\right)$. It is evident that the new-physics effects do not decouple neither from the beta function nor from the polarization function $\Pi^{(4+n){\rm D}}_{\rm KK}(p^2)$, as it occurs in the case of one extra dimension.\\

\noindent \textit{Mass-dependent scheme}. The structure of the new counterterms suggests to use the following renormalization conditions:
\begin{subequations}
\begin{align}
&\Pi^{(4+n){\rm D}}_{\rm KK}(p^2=-\mu^2)=0\, , \\
&\frac{d}{dp^2}\Pi^{(4+n){\rm D}}_{\rm KK}(p^2)|_{p^2=-\mu^2}=0\, , \\
&\vdots  \nonumber \\
&\frac{d^{[\frac{n}{2}]}}{d(p^2)^{[\frac{n}{2}]}}\Pi^{(4+n){\rm D}}_{\rm KK}(p^2)|_{p^2=-\mu^2}=0\, .
\end{align}
\end{subequations}
From these renormalization conditions, the derivation of the counterterms for relatively large values of $n$ does not seem to be a simple task. Fortunately, the symmetry of the theory allows one to write closed expressions for any value of $n$. The counterterms induced by interactions of canonical dimension higher than four, can be determined, for any value of $n$, from the expression
\begin{eqnarray}
\label{CTK}
\delta^{\left([\frac{n}{2}]-m\right)}_A&=&-\lambda_{\left([\frac{n}{2}]-m\right)}-A_{\left([\frac{n}{2}]-m\right)}(n,\epsilon)\nonumber \\
&&-\sum^m_{k=1}\frac{(-1)^k{\left([\frac{n}{2}]-m+k\right)}!}{k!{\left([\frac{n}{2}]-m\right)}!}\left[\lambda_{\left([\frac{n}{2}]-m+k\right)}+
\delta^{\left([\frac{n}{2}]-m+k\right)}_A+A_{\left([\frac{n}{2}]-m+k\right)}(n,\epsilon)\right]\left(\frac{\mu^2}{R^{-2}}\right)\nonumber \\
&&-\frac{g^2N}{(4\pi)^2}\int^1_0dx\left\{
\left[f(x)\left(\frac{R^{-2}}{\mu^2}\right)^{[\frac{n}{2}]-m}+
h(x)\left[-x(1-x)\right]^{[\frac{n}{2}]-m}\frac{d^{[\frac{n}{2}]-m}F(n,\bar{c}^2)}{d(\bar{c}^2)^{[\frac{n}{2}]-m}}\right]\right\}\, ,
\end{eqnarray}
where the definitions
\begin{subequations}
\begin{align}
h(x)=&f(x)-\frac{n}{2}g(x)\, , \\
A_{(k)}(n,\epsilon)=&\frac{g^2N}{(4\pi)^2}\int^1_0dx\,h(x)\left[f_{(k)}(n)\frac{1}{\sqrt{\pi}}\Gamma\left(\frac{1+\epsilon}{2}\right)\zeta(1+\epsilon)+
g_{(k)}(n)\Gamma\left(\frac{\epsilon}{2}\right)\zeta(\epsilon)+F_{(k)}(n)\right][-x(1-x)]^k
\end{align}
\end{subequations}
have been introduced. Notice that the counterterms in (\ref{CTK}) are divergent through the quantity $A_{(k)}(n,\epsilon)$, which contains divergences arising from short-distance effects in the compact manifold. Also note that, in expression (\ref{CTK}), $m$ goes from 0 to $[\frac{n}{2}]-1$. So for $m=0$, $\delta^{[\frac{n}{2}]}_A$ is determined in terms of $\lambda_{[\frac{n}{2}]}$ and $A_{[\frac{n}{2}]}(n,\epsilon)$. This result allows us to determine $\delta^{[\frac{n}{2}]-1}_A$. In turn,  $\delta^{[\frac{n}{2}]}_A$ and $\delta^{[\frac{n}{2}]-1}_A$ allow us to determine $\delta^{[\frac{n}{2}]-3}_A$, and so on.\\

As far as the usual counterterm is concerned, it can be derived, for any value of $n$, from the following expression:
\begin{eqnarray}
\label{CTU}
\delta_A&=&\frac{g^2N}{(4\pi)^2}\int^1_0 \Bigg\{f(x)\left[\Gamma\left(\frac{\epsilon}{2}\right)\left(\frac{\bar{\Delta}^2}{4\pi\hat{\mu}^2}\right)^{-\frac{\epsilon}{2}} +\left[\frac{n}{2}\right] \right]\nonumber \\
&&+h(x)\left[A_{(0)}(n,\epsilon)+\sum^{[\frac{n}{2}]}_{k=0}(-1)^k (\bar{c}^2)^k\frac{d^kF(n,\bar{c}^2)}{d(\bar{c}^2)^k}\right] \Bigg\}\, ,
\end{eqnarray}
where
\begin{equation}
A_{(0)}(n,\epsilon)=g_{(0)}(n)\left(\frac{R^{-2}}{4\pi \hat{\mu}^2}\right)^{-\frac{\epsilon}{2}}\Gamma\left(\frac{\epsilon}{2}\right)\zeta(\epsilon)+F_{(0)}(n)\, .
\end{equation}
This counterterm contains only divergences that arise from short-distance effects in the usual spacetime manifold. The ultraviolet divergence of the zero mode appears in the first term of (\ref{CTU}), proportional to $f(x)$, while the ultraviolet divergences induced by the KK excited modes are contained in the factor $A_{(0)}(n,\epsilon)$. \\

On the other hand, the renormalized polarization function is given by
\begin{eqnarray}
\Pi^{(4+n){\rm D}}_{\rm KK}(p^2)&=&\frac{g^2N}{(4\pi)^2}\int^1_0 dx\Bigg\{f(x)\left[\log\left(\frac{\bar{\Delta}^2_{(\underline{0})}}{\Delta^2_{(\underline{0})}}\right) -\left[\frac{n}{2}\right] \right]\nonumber \\
&&+h(x)\left[F(n,c^2)-F(n,\bar{c}^2)-\sum^{[\frac{n}{2}]}_{k=1}\left(c^2-\bar{c}^2\right)^k\frac{d^kF(n, \bar{c}^2)}{d(\bar{c}^2)^k}  \right] \Bigg\}\nonumber \\
&&+\sum^{[\frac{n}{2}]}_{k=1}\left[\frac{1}{k!}\frac{d^k \delta^{(k)}}{d\left(\frac{R^{-2}}{\mu^2}\right)^k}\right]\left(\frac{p^2}{\mu^2}\right)^k\, .
\end{eqnarray}

We now proceed to derive the beta function. Of course, there is a beta function for each coupling constant, but we will focus on the usual beta function, which emerges from the standard countererm (\ref{CTU}). The usual beta function does not depend on the scale, so the $(4+n)$-dimensional beta function can be written as follows:
\begin{equation}
\label{beta1}
\beta_{(4+n){\rm D}}(g)=\beta(g)+\frac{g^2N}{(4\pi)^2}\int^1_0dx\, h(x)\, \bar{c}^2\frac{d}{d\bar{c}^2}\left[\sum^{[\frac{n}{2}]}_{p=0}(-1)^p(\bar{c}^2)^p\frac{d^pF(n,\bar{c}^2)}{d(\bar{c}^2)^p}\right]\, ,
\end{equation}
where the term $p=0$, within square brackets, corresponds to $F(n,\bar{c}^2)$. To investigate the impact of extra dimensions on beta function, we need to determine the sign of the new-physics contribution. The calculation of the derivatives on the $F(n,\bar{c}^2)$ function is straightforward. In addition, both the finite sum on the index $p$ and the parametric integral on the variable $x$ can be performed. Once this is done, we get
\begin{eqnarray}
\label{beta2}
\beta_{(4+n){\rm D}}(g)&=&\beta(g)+\frac{g^2N}{(4\pi)^2}\frac{1}{2^{n-1}}\sum^n_{r=1}\sum^{r}_{l=1}\left(\begin{array}{ccc}
n \\
l-1
\end{array}\right)\pi^{\frac{n-r}{2}}\nonumber \\
&\times& \sum^{\infty}_{k=[\frac{n}{2}]+1}I(n,k)S(n,k)\Gamma\left(\frac{2k+r-n}{2}\right)\zeta(2k+r-n)
\left(\frac{\mu^2}{R^{-2}}\right)^k\, .
\end{eqnarray}
In this expression, $S(n,k)$ is the finite sum over the $p$ index, which is given by:
\begin{equation}
S(n,k)=\frac{1}{\mathbf{e}}\left[\frac{\Gamma(k+1,-1)}{\Gamma(k+1)}
-\frac{\Gamma\left(k-[\frac{n}{2}],-1\right)}{\Gamma\left(k-[\frac{n}{2}]\right)}\right]\, ,
\end{equation}
where  $\Gamma(a,z)$ is the incomplete gamma function and $\mathbf{e}$ is the Euler-Napier constant. On the other hand, $I(n,k)$ is the parametric integral over the $x$ variable, which can be expressed as:
\begin{eqnarray}
I(n,k)&=&\int^1_0 dx h(x)\left[x(1-x)\right]^k \nonumber \\
&=&\frac{\sqrt{\pi}}{2^{2k+1}}\frac{\Gamma(k+1)(16k+22+n)}{(2k+1)(2k+3)\Gamma\left(k+\frac{1}{2}\right)}\, .
\end{eqnarray}
Note that $I(n,k)>0$ for any value of $n$ and $k$. In fact, all the quantities that appear in Eq.(\ref{beta2}) are positive, with the exception of the sum $S(n,k)$, which is negative for $k = [\frac{n}{2}]+1$, positive  for $k = [\frac{n}{2}]+2$, negative for $k = [\frac{n}{2}]+3$, and so on. However, the contribution to the beta function decreases considerably, in absolute value, for increasing values of the $k$ index. Then, the contribution $ k = [\frac {n} {2}] + 1 $ dominates the contributions coming from the other terms, which alternate positive and negative signs, but are insignificant in absolute value. Keeping only the dominant contribution, Eq. (\ref{beta2}) becomes
\begin{eqnarray}
\label{beta3}
\beta_{(4+n){\rm D}}(g)&=&\beta(g)+\frac{g^2N}{(4\pi)^2}\frac{1}{2^{n-1}}\sum^n_{r=1}\sum^{r}_{l=1}\left(\begin{array}{ccc}
n \\
l-1
\end{array}\right)\pi^{\frac{n-r}{2}}I\left(n,[\frac{n}{2}]+1\right)S\left(n,[\frac{n}{2}]+1\right)\nonumber \\
&\times&\Gamma\left(\frac{2[\frac{n}{2}]+r-n+2}{2}\right)\zeta\left(2[\frac{n}{2}]+r-n+2\right)
\left(\frac{\mu^2}{R^{-2}}\right)^{[\frac{n}{2}]+1}+\cdots\, ,
\end{eqnarray}
where the ellipsis denote subdominant terms. In this case,
\begin{equation}
S\left(n,[\frac{n}{2}]+1\right)=-1+\frac{1}{\mathbf{e}}\frac{\Gamma\left([\frac{n}{2}]+2,-1\right)}{\Gamma\left([\frac{n}{2}]+2\right)}< 0\, ,
\end{equation}
which means that $\beta_{(4+n){\rm D}}(g)<0$. \\

The above results show that the radiative correction induced by the KK excitations has the same sign as that of the usual theory, which means that the KK excitations behave like genuine Yang-Mills fields. The new physics effect decouples, since $\beta_{(4+n){\rm D}}(g) \to \beta(g)$ when $R^{-1}$ goes to infinity. This is the main result of the present work.

\section{Concluding remarks and summary}
\label{C} In the present paper, we have shown that Yang-Mills theories in more than four universal extra dimensions remain perturbative at the one loop level. We have focused  the problem not from the very intuitive approach of using a cutoff regulator, but from the perspective of the dimensional regularization scheme. In this approach, large contributions of continuous and discrete momenta are kept, but they are removed by adjusting parameters of appropriate counterterms, which are already available, since the effective Lagrangian contains all interactions that respect the symmetries of the theory.\\

A comprehensive study of the beta has been presented. The effective KK theory was quantized using a ${\rm SU}(N,{\cal M}^4)$-covariant gauge-fixing procedure, which considerably simplifies the calculations. This class of theories, which are characterized by an infinite number of fields, generates loop amplitudes that involve discrete and continuous sums $\sum_{(\underline{k})}\int d^4 k$, each of which may diverge for large discrete momentum $k_{\bar \mu}$ or for large continuous momentum $k_\mu$. We regularized both types of potentially divergent sums by using the dimensional-regularization approach, which allowed us to parameterize the one-loop amplitudes as products $\Gamma\left(\frac{\epsilon}{2}\right) E^{c^2}_l\left(\frac{\epsilon}{2}\right)$, with $E^{c^2}_l(s)$ the $l$-dimensional Epstein zeta function representing the regularized discrete sums. In our case, the multidimensional Epstein functions depend on $c^2\propto \frac{p^2}{R^{-2}}$, with $p^2$ the external momentum. It was argued that divergences arising as poles of the Epstein function correspond to genuine ultraviolet divergences, since they are associated with large values of discrete momenta $k_{\bar \mu}$ or, equivalently, with short-distance effects in the compact manifold. Thus, this type of divergences can be removed through renormalization, just as it is done in the case of ultraviolet divergences emerging from short-distance effects in the usual four-dimensional spacetime manifold. Results available in the literature were used to reduce $E^{c^2}_l(s)$ into a finite sum of one-dimensional Epstein functions. This result plays a relevant role because it allows us to distinguish the two types of ultraviolet divergences already commented. To see this, note that
\begin{equation}
\Gamma\left(\frac{\epsilon}{2}\right) E^{c^2}_l\left(\frac{\epsilon}{2}\right)=\Gamma\left(\frac{\epsilon}{2}\right) \left[E^{c^2}_1\left(\frac{\epsilon}{2}\right) + \frac{(-1)^{l-1}}{2^{l-1}}\sum^{l-1}_{p=1}\left(\begin{array}{ccc}
l-1 \\
p
\end{array}\right)(-1)^p \pi^{\frac{p}{2}}\frac{\Gamma\left(\frac{\epsilon}{2}-\frac{p}{2}\right)}
{\Gamma\left(\frac{\epsilon}{2}\right)}E^{c^2}_1\left(\frac{\epsilon}{2}-\frac{p}{2}\right)\right]\, .
\end{equation}
The first term of this expression represents the usual ultraviolet divergences induced by the KK excited modes, since the Epstein function $E^{c^2}_1\left(\frac{\epsilon}{2}\right)$ converges for $\epsilon \to 0$ and does not depend on $c^2$. The divergences arise exclusively from the pole of the gamma function. The other type of divergences arises from the second term of the above expression. Since the factor $\Gamma\left(\frac{\epsilon}{2}\right)$ is canceled by the denominator of the sum over index $p$, divergences arise from products $\Gamma\left(\frac{\epsilon}{2}-\frac{p}{2}\right) E^{c^2}_1\left(\frac{\epsilon}{2}-\frac{p}{2}\right)$ of this sum when $\epsilon \to 0$. It can be appreciated from this expression that for even values of $p$ divergences arise as poles of the gamma function when $\epsilon \to 0$, while for odd values of $p$ divergences emerge as poles of the one-dimensional Epstein function in this limit, since $E^{c^2}_1(s)$ has poles at $s=\frac{1}{2}, -\frac{1}{2}, -\frac{3}{2},\cdots$. This is the type of ultraviolet divergences which we have identified as short-distance effects in the compact manifold. However, note that this type of divergences does not exist in the five-dimensional theory, since in this case $l=1$. Because this divergent Epstein function depends on power of the external momentum, it only can be associated with interactions of canonical dimension higher than four. Using gauge invariance as a guide, we have introduced all interactions needed to generate the required counterterms, which are already available since the effective theory contains all the interactions that respect the symmetries of the theory. To carry out practical calculations, we have used another important result of the literature that allows us to express the one-dimensional Epstein function in terms of products of gamma and Riemman functions through a power series in $c^2$. This expansion in powers of $c^2$ is valid in our case because this effective theory aims at the description of physics at energies lower than the compactification scale $R^{-1}$. At this level, the poles of the $\Gamma\left(\frac{\epsilon}{2}-\frac{p}{2}\right) E^{c^2}_1\left(\frac{\epsilon}{2}-\frac{p}{2}\right)$ products translate into poles of both the Riemann and gamma functions. This decomposition of the Epstein function has a high practical value, since both the gamma function and the Riemann zeta function already appear in libraries of modern technical computing systems, such as Mathematica. To be confident about our results, we have systematically used this program for the present investigation.\\

The case of one extra dimension was studied separately. Our motivation to do this is that this theory has some peculiarities that make it substantially different from the more general case of an arbitrary number of extra dimensions. An interesting property of the five-dimensional formulation is that no divergences associated with the one-dimensional Epstein function emerge, so the only divergences are the usual ultraviolet divergences generated by the KK zero- and excited-modes. This means that the usual counterterm
$-\frac{\delta_A}{4}F^{(\underline{0})a}_{\mu \nu}F^{(\underline{0})\,\mu \nu}_a$ (in our gauge-invariant quantization scheme) is enough to remove  divergences. Another important property of this case is that all Green's functions of canonical dimension higher than four in which external legs are all zero-modes are free of ultraviolet divergences as long as no interactions of dimension higher than four are inserted in the loops. This has been verified in phenomenological applications.\\

The beta function was studied in both the five-dimensional and the $(4+n)$-dimensional cases. The analysis was performed in a mass-independent scheme and in a mass-dependent scheme as well. In the former scheme, it was found that neither the beta function nor the renormalized polarization function exhibits decoupling of extra dimensions effects, while in the latter scheme the decoupling is manifest. In the $(4+n)$-dimensional case, explicit expressions for counterterms, renormalized polarization function, and beta function, valid for any value of $n$, were presented. It was shown that the excited KK modes induce a radiative correction that reinforces the value of the usual beta function, since it has the same sign. On physical grounds, this result was expected, since the KK excitations of the YM zero mode are also gauge fields governed by the non-Abelian symmetry. However, we have seen that proving this fact is not a trivial task, mainly because dimensional regularization must be used along with an appropriate renormalization scheme.\\

The systematic use of multidimensional zeta functions that arise from infinite discrete sums regularized in the dimensional-regularization scheme, played a central role in our study. Our conclusion is that short-distance effects of zero mode and excited modes in the spacetime manifold arise as poles of the gamma function when $\epsilon=4-D$ approaches zero, while short-distance effects of excited modes in the compact manifold emerge as poles of the one-dimensional Epstein function.\\

We emphasize that our identification of two types of ultraviolet divergences turns out to be natural from the perspective that we have two different spaces, one infinite and the other compact. We think that this fact, along with our approach to this type of effective theories, constitutes an important contribution of this work.\\

To finish, we would like to comment that our method for handling divergences that arise (for $n\geq 2$ ) from  poles of the Epstein function could be very useful in phenomenological studies of universal extra dimensions. Especially in some physical processes that are naturally suppressed in the Standard Model, so they can be very  sensitive to new-physics effects. This is the case of physical processes that first arise at one loop, as rare Higgs boson decays or flavor violation, which are free of short-distance effects in the usual manifold, but not in the compact manifold for $n\geq2$.\\

\acknowledgments{We acknowledge financial support from Consejo Nacional de Ciencia y Tecnolog\'\i a (CONACYT). H.N.S. and J.J.T. also acknowledge financial support from CONACYT through Sistema Nacional de Investigadores (SNI).}

\section*{References}


\end{document}